\documentclass[onecolumn,floatfix,showpacs,aps]{revtex4}
\usepackage{amsfonts,amsmath,amssymb,amsthm}
\usepackage{graphicx,epsfig} 
 
\begin{document}
\title{Fault-tolerant quantum computation versus Gaussian noise}

\author{Hui Khoon Ng and John Preskill}

\affiliation{Institute for Quantum Information, California Institute of Technology, Pasadena, CA 91125}

\begin{abstract}

We study the robustness of a fault-tolerant quantum computer subject to Gaussian non-Markovian quantum noise, and we show that scalable quantum computation is possible if the noise power spectrum satisfies an appropriate ``threshold condition.'' Our condition is less sensitive to very-high-frequency noise than previously derived threshold conditions for non-Markovian noise. 
\pacs{03.67.Pp, 03.67.Lx}
\end{abstract}

\maketitle
\section{Introduction}
\label{sec:introduction}

The theory of fault-tolerant quantum computation shows that properly encoded quantum information can be protected against decoherence and processed reliably with imperfect hardware \cite{Shor96}. Demonstrating that this theory really works in practice is one of the great challenges facing contemporary science. A large-scale fault-tolerant quantum computer would be a scientific milestone, and it should also be {\em useful}, capable of solving hard problems that are beyond the reach of ordinary digital computers.

Though the theory of quantum fault tolerance strengthens our confidence that truly scalable quantum computers can be realized in the next few decades, failure is certainly possible. Perhaps the engineering challenges will prove to be so daunting, and the resources needed to overcome them so demanding, that society will be unable or unwilling to bear the cost for the foreseeable future. Perhaps new fundamental principles of physics, as yet undiscovered, will prevent large-scale quantum computers from behaving as currently accepted theory dictates. Finding that quantum computers fail for a fundamental reason would be a significant scientific advance, but would disappoint prospective users.

There is a third reason to worry about the future prospects for fault-tolerant quantum computing. Mathematical results establishing that fault tolerance works effectively are premised on assumptions about the properties of the noise. The most obvious requirement is that the noise must be sufficiently weak --- if the noise strength is below a {\em threshold of accuracy} then quantum computing is scalable in principle. But in addition, the noise must be suitably local, both spatially and temporally. Perhaps the quest for a quantum computer will be frustrated because the noise afflicting actual hardware is just not amenable to fault-tolerant protocols. 

We can anticipate therefore that progress toward scalable quantum computing will require an ongoing dialog between experimenters who will better understand the limitations of their devices and theorists who will propose better ways to overcome the limitations and to evaluate the efficacy of these proposals. In the meantime, an important task for theorists is to broaden the range of noise models for which useful accuracy threshold theorems can be proven, and we pursue that task in this paper. Our main result is a new proof of the threshold theorem for non-Markovian Gaussian noise models, in which system qubits are locally coupled to bath variables that have Gaussian fluctuations. Specifically, if the bath is a system of uncoupled harmonic oscillators, at either zero or nonzero temperature, our theorem expresses the threshold condition in terms of the power spectrum of the bath fluctuations.

Early proofs of the threshold theorem \cite{ben-or,kitaev-threshold,knill-laflamme} assumed that the noise is {\em Markovian}. This means that each quantum gate in the noisy circuit can be modeled as a unitary transformation that acts jointly on a set of the qubits in the computer (the system qubits) and on the environment (the bath variables), but where it is assumed that the bath has no memory --- the state of the bath is refreshed after every gate. The theorem was extended to a class of non-Markovian noise models in \cite{terhal}, and further generalized in \cite{AGP} and \cite{AKP}. The results of \cite{AGP,AKP} have the substantial virtue that the state of the bath and its internal dynamics can be arbitrary; for fault-tolerant quantum computing to work, it is only required that the bath couple weakly and locally to the system. 

However these results also have two serious drawbacks. First, the threshold condition is not easily related to experimentally accessible quantities; rather it requires terms in the Hamiltonian that couple the system to the bath to have a sufficiently small operator norm. Second, this condition severely constrains the very-high-frequency fluctuations of the bath. Intuitively, it seems that this constraint, which may limit the applicability of the threshold theorem to noise in some realistic settings, ought not to be necessary, since fluctuations with a time scale much shorter than the time it takes to execute a quantum gate tend to average out.

One possible way to reach more pleasing conclusions is to make physically reasonable assumptions about the noise that go beyond the assumptions of \cite{AGP,AKP}; that is the approach we follow here. Our new threshold theorem applies to any noise model in which the bath variables are free fields (aside from their coupling to the system qubits), and expresses the threshold condition in terms of the bath's two-point correlation function, which is in principle measurable. It should be possible to extend our analysis to the case where the bath variables have sufficiently weak self-interactions, though we will not pursue that extension here. Furthermore, though our new threshold condition still requires the very-high-frequency bath fluctuations to be sufficiently weak, this requirement is considerably relaxed compared to previous threshold theorems that apply to non-Markovian noise. Showing that these requirements can be relaxed even further, perhaps by making additional physically motivated assumptions, is an important open problem. 

Experimenters use a variety of techniques to suppress the noise in quantum hardware, such as cleverly designed pulse sequences to improve the fidelity of quantum gates (spin echos, dynamical decoupling, etc.) and intrinsically robust encodings of quantum information (noiseless subsystems, topologically protected qubits, etc.). These techniques can be highly effective and are likely to be incorporated into the design of future quantum computers, but do not by themselves suffice to ensure the scalability of quantum computing. After such tricks are exhausted some residual noise inevitably remains that must be controlled using quantum error-correcting codes and fault-tolerant methods. Since our objective in this paper is to study the effectiveness of these fault-tolerant methods, our noise models may be viewed as effective descriptions of this residual noise in ``fundamental'' quantum gates that might already be realized using complex and sophisticated protocols.

After reviewing previously known formulations of the quantum accuracy threshold theorem in Sec.~\ref{sec:models} (with some details relegated to Appendix A), we state our new result in Sec.~\ref{sec:gaussian}, explore some of its implications in Sec.~\ref{sec:implications}, derive it in Sec.~\ref{sec:derivation}, and discuss some generalizations in Sec.~\ref{sec:generalizations}. We derive a sharper result for the case of pure dephasing noise in Sec.~\ref{sec:diagonal}. Sec.~\ref{sec:conclusions} contains our conclusions. 

\section{Noise models and quantum accuracy threshold theorems}
\label{sec:models}

Here we will briefly review some previously know formulations of the quantum accuracy threshold theorem, and explain why these results still leave something to be desired. Then in Sec.~\ref{sec:gaussian} we will state our new result, which addresses some of the shortcomings of the previous results. 

The goal of fault-tolerant quantum computing is to simulate an ideal quantum circuit using the noisy gates that can be executed by actual devices. Theoretical results show that this goal is attainable if the noise is not too strong and not too strongly correlated. The essential trick that makes fault tolerance work is that the logical quantum state processed by the computer can be encoded very nonlocally, so that it is well protected from damage caused by local noise.

It is convenient to analyze the effectiveness of a fault-tolerant noisy circuit by invoking a {\em fault-path expansion}; schematically,
\begin{equation}
\label{fault-path-expansion}
{\rm Noisy ~ Circuit} = \sum {\rm ``Fault ~ Path"}~.
\end{equation}
Let us use the term {\em location} to speak of an operation in a quantum circuit that is performed in a single time step; a location may be a single-qubit or multi-qubit gate, a qubit preparation step, a qubit measurement, or the identity operation in the case of a qubit that is idle during the time step. In each fault path, the quantum gates are faulty at a specified set of locations in the circuit, while at all other locations the quantum gates are assumed to be ideal. We say that the faulty locations are ``bad'' and that the ideal locations are ``good.'' The general concept of a fault-path expansion applies quite broadly, and different noise models can be distinguished according to how we flesh out the meaning of eq.~(\ref{fault-path-expansion}).

\subsection{Local stochastic noise}
\label{local-stochastic-noise}

In a ``stochastic'' noise model we assign a {\em probability} to each fault path \cite{AGP}. We speak of {\em local stochastic noise} with strength $\varepsilon$ if, for any specified set $\mathcal{I}_r$ of $r$ locations in the circuit, the sum ${P}^{\rm bad}(\mathcal{I}_r)$ of the probabilities of all fault paths that are bad at {\em all} of these $r$ locations satisfies
\begin{equation}
{P}^{\rm bad}(\mathcal{I}_r) \le \varepsilon^r~.
\end{equation}
In this noise model, no further restrictions are imposed on the noise, and in particular the trace-preserving quantum operation applied at the faulty locations may be chosen for each fault path by an adversary who wants the computation to fail. Thus the faults can be correlated, both spatially and temporally, but the adversary's power is limited because an attack on $r$ specified circuit locations occurs with probability at most $\varepsilon^r$. The noise is ``local'' in the sense that attacking each additional location suppresses the probability of the fault path by another power of $\varepsilon$.

Most proofs of the threshold theorem use {\em recursive simulations}. This means that quantum information is protected by a hierarchy of codes within codes, and that the fault-tolerant circuit has a self-similar structure. We refer to an unencoded quantum circuit as a ``level-0'' simulation. In a level-1 simulation, each elementary gate in the level-0 circuit is replaced by a level-1 {\em gadget} constructed from elementary gates; this 1-gadget performs the appropriate encoded operation on logical qubits that are protected by a quantum error-correcting code $\mathcal{C}$. In a level-2 simulation, each elementary gate in the ideal circuit is replaced by a level-2 gadget; the 2-gadget is constructed by replacing each elementary gate in the 1-gadget by a 1-gadget. A 2-gadget operates on quantum information protected by $\mathcal{C}\triangleright\mathcal{C}$, where $\triangleright$ denotes code concatenation. (That is, $\mathcal{C}_1\triangleright\mathcal{C}_2$ is encoded by first encoding the ``outer'' code $\mathcal{C}_2$, and then encoding each qubit in the $\mathcal{C}_2$ block using the ``inner'' code $\mathcal{C}_1$.) In a level-$k$ simulation, each elementary gate in the ideal circuit is replaced by a level-$k$ gadget, constructed by replacing each elementary gate in the $(k{-}1)$-gadget by a 1-gadget; it operates on quantum information protected by $\mathcal{C}^{\triangleright k}$. 

For local stochastic noise, and also for other noise models with suitable properties, a recursive simulation can be analyzed by a procedure called {\em level reduction}, in which a level-$k$ simulation is mapped to a ``coarse-grained'' level-$(k{-}1)$ simulation that acts on the top-level logical information in exactly the same way. Suppose for example that $\mathcal{C}$ is a distance-3 code that can correct one error. Then if the 1-gadgets are properly designed, each ``good'' 1-gadget that contains no more than one faulty location simulates the corresponding ideal gate correctly, while ``bad'' 1-gadgets with more than one fault may simulate the ideal gate incorrectly. In the level reduction step, for each fault path the good 1-gadgets are mapped to ideal level-0 gates, while the bad 1-gadgets are mapped to faulty level-0 gates. After this step, the resulting noisy circuit is still subject to local stochastic noise, but with a renormalized value of the noise strength
\begin{equation}
\label{noise-renormalization}
\varepsilon^{(1)} = \varepsilon^2/\varepsilon_0 = \varepsilon_0\left(\varepsilon/\varepsilon_0\right)^2~.
\end{equation}
The renormalized value of the noise strength is $O(\varepsilon^2)$, because at least two faults are required for a 1-gadget to fail; the quantity $\varepsilon_0^{-1}$ is a combinatoric factor counting the number of ``malignant'' sets of locations within the 1-gadget where faults can cause failure. 

Since level reduction maps local stochastic noise to local stochastic noise (but with a revised value of the noise strength), the level reduction step can be carried out repeatedly, and analyzed by the same method each time. That the structure of the noise is preserved, even though its strength is renormalized, is a useful feature of the local stochastic noise model not shared by some noise models. For example, if faults in level-0 gates were independently and identically distributed, the effective noise model after one level reduction step would become correlated rather than independent. See \cite{AGP,Aliferis07} for a more detailed discussion of the level reduction procedure. 

By repeating the level reduction step all together $k$ times, we reduce the level-$k$ simulation to an effective level-0 ({\em i.e.}, unencoded) simulation with noise strength
\begin{equation}
\label{level-k-strength}
\varepsilon^{(k)} = \varepsilon_0\left(\varepsilon/\varepsilon_0\right)^{2^k}~.
\end{equation}
It follows that for $\varepsilon < \varepsilon_0$ (the {\em accuracy threshold}), the effective noise strength becomes negligibly small for $k$ sufficiently large, and the simulation becomes highly reliable. More precisely, for any fixed $\varepsilon < \varepsilon_0$ and fixed $\delta > 0$, an ideal circuit with $L$ gates can be simulated with error probability $\delta$ by a noisy circuit with $L^*$ gates, where for some constant $c$
\begin{equation}
L^*/L =O\left(\left( \frac{\log (L/\delta)}{\log(\varepsilon_0/\varepsilon)} \right)^c \right)~
\end{equation}
(The constant $c$ is determined by the size of the 1-gadgets.) Thus, with reasonable overhead cost, the noisy simulation gets the right answer with high probability.  This is the quantum accuracy threshold theorem for local stochastic noise.

For the threshold theorem to apply, two features of the simulation are essential: First, we must assume that quantum gates can be executed in parallel --- otherwise we would be unable to control storage errors that occur simultaneously in different parts of the computer. Second, we assume that qubits can be ``discarded'' and replaced by fresh qubits (for example, by measuring the qubits and resetting them) --- otherwise we would be unable to flush from the computer the entropy introduced by noise. Estimates of the accuracy threshold $\varepsilon_0$ often rely on further assumptions. For example, if we assume that qubit measurements are as fast as quantum gates, that classical computations are arbitrarily accurate, that the accuracy of a two-qubit quantum gate does not depend on the spatial separation of the qubits, and that no data qubits ``leak'' from the computational Hilbert space, then it has been shown that $\varepsilon_0 > .67 \times 10^{-3}$ \cite{fibonacci}. For noise models with weaker correlations than in the local stochastic noise model, the proven accuracy threshold is above $10^{-3}$ \cite{AGP2,Reichardt06}, and numerical evidence suggests that the actual value of the threshold can be of order $1\%$ \cite{knill,raussendorf}. Furthermore, it has been shown that the threshold is not drastically reduced if some of these assumptions are relaxed, for example by allowing measurements to be slow \cite{slow}, allowing leakage \cite{leakage}, or requiring quantum gates to be local on a two-dimensional array \cite{svore}. 

\subsection{Local non-Markovian noise}
\label{subsec:local-noise}

The local stochastic noise model is handy for analysis and has some quasi-realistic features, but it is still rather artificial. From a physics perspective, it is more natural to formulate the noise model in terms of a Hamiltonian $H$ that governs the joint evolution of the system and the bath. We may express $H$ as
\begin{equation}
H=H_S+H_B +H_{SB}~,
\end{equation}
where $H_S$ is the time-dependent system Hamiltonian that realizes the ideal quantum circuit, $H_B$ is the (arbitrary) Hamiltonian of the bath, and  $H_{SB}$ is a perturbation, responsible for the noise, that may couple the system to the bath. We say that such a noise model is {\em non-Markovian}, meaning that quantum information can escape from the system to the bath and then return to the system at a later time, so that the state of the system at time $t + dt$ is not uniquely determined by its state at time $t$. Furthermore, $H_{SB}$ may also contain terms that act nontrivially only on the system, representing unitary noise arising from imperfect control of the system Hamiltonian. Actually, the local stochastic noise model already incorporates some non-Markovian effects; even when fault paths are weighted by probabilities, the adversary who attacks the circuit might employ a quantum memory. But different methods are needed to analyze the consequences of Hamiltonian noise models, because fault paths are summed coherently rather than stochastically.

The locations in a quantum circuit include not only quantum gates and storage steps, but also qubit preparation and measurement steps. Preparation and measurement noise can be incorporated into a Hamiltonian description by various means. In this paper we will take an especially simple approach, modeling an imperfect preparation by an ideal preparation followed by evolution governed by $H$, and modeling an imperfect measurement by an ideal measurement preceded by evolution governed by $H$. For the time being, to simplify the discussion, we will imagine that system qubits are prepared only at the very beginning of the computation and measured only at the very end. Preparations and measurements that occur at intermediate times can easily be incorporated; we will elaborate on this point in Sec.~\ref{sec:generalizations}. For the continuous-time Hamiltonian dynamics we are now considering, a ``location'' consists of a specified qubit or set of qubits to which a gate is applied, and a specified {\em time interval} during which that gate is realized by the ideal system Hamiltonian $H_S$.

We may say that the Hamiltonian noise model is ``local'' if the perturbation $H_{SB}$ can be expressed as a sum of terms
\begin{equation}
\label{expansion-system-bath-Hamiltonian}
H_{SB}= \sum_a H_{SB}^{(a)}~,
\end{equation}
where each $H_{SB}^{(a)}$ acts on only a small number of system qubits (while perhaps also acting collectively on many bath variables). The joint unitary time evolution operator $U_{SB}$ for system and bath, resulting from integrating the Schr\"odinger equation for Hamiltonian $H$, can be formally expanded to all orders in time-dependent perturbation theory in $H_{SB}$. In any fixed term in this expansion, perturbations chosen from the set $\{H_{SB}^{(a)}\}$ are inserted at specified times. For such a fixed term in the perturbation expansion, let us say that a location in the (level-0) noisy simulation is ``bad'' if an inserted perturbation acts nontrivially somewhere inside that location; otherwise that location is ``good.'' Of course, under this definition a single insertion of $H_{SB}^{(a)}$ might cause two (or perhaps more) locations to be bad in a particular time step, if $H_{SB}^{(a)}$ acts collectively on two qubits that are undergoing different gates executed in parallel in the ideal circuit.

As already noted, we may assume that the system qubits have been initialized ideally at the start of the Hamiltonian evolution; we denote this initial system state by $|\Psi_S^0\rangle$. We also assume that the initial state of the bath is a pure state $|\Psi_B^0\rangle$. There is really no loss of generality in supposing that the bath starts out in a pure state; if we wish to consider a mixed initial state of the bath instead (for example, a thermal state), we may include in the bath a ``reference'' system that ``purifies'' the mixed state.

We have also noted that we may assume that final measurements performed on system qubits are ideal. Just before these final measurements are conducted, the (pure) state of the system and bath is
\begin{equation}
|\Psi_{SB}\rangle=U_{SB}|\Psi_{SB}^0\rangle~,
\end{equation}
where $|\Psi_{SB}^0\rangle = |\Psi_S^0\rangle\otimes |\Psi_B^0\rangle$ is the initial state of system and bath. For any specified set $\mathcal{I}_r$ of $r$ locations in the circuit,  let us denote by $|\Psi_{SB}^{\rm bad}(\mathcal{I}_r)\rangle$ the sum of all the terms in the formal perturbation expansion of $|\Psi_{SB}\rangle$ such that all of these $r$ locations are bad. Then we speak of {\em local non-Markovian noise} (or more briefly {\em local noise}) with strength $\varepsilon$ if
\begin{equation}
\label{local-noise-strength}
\| |\Psi_{SB}^{\rm bad}(\mathcal{I}_r)\rangle\| \le \varepsilon^r~.
\end{equation}
The noise strength $\varepsilon$ can be related to properties of the perturbation $H_{SB}$. We will sometimes refer to this model as local {\em coherent} noise, to emphasize that (in contrast to the local {\em stochastic} noise model) fault paths are assigned amplitudes rather than probabilities.

Although there are some new subtleties (see \cite{AGP} and Appendix~\ref{sec:appendix}), the level-reduction concept can be applied to Hamiltonian noise models in much the same way as for stochastic models. We may say that a 1-gadget is bad if it contains bad level-0 gates at a malignant set of locations, that a 2-gadget is bad if it contains bad 1-gadgets at a malignant set of locations, that a 3-gadget is bad if it contains bad 2-gadgets at a malignant set of locations, and so on. For any specified set $\mathcal{I}_r^{(k)}$ of $r$ $k$-gadgets in the circuit,  let us denote by $|\Psi_{SB}^{\rm bad}(\mathcal{I}_r^{(k)})\rangle$ the sum of all the terms in the formal perturbation expansion of $|\Psi_{SB}\rangle$ such that all of these $r$ $k$-gadgets are bad. Then it follows from eq.~(\ref{local-noise-strength}) that 
\begin{equation}
\label{local-noise-strength-levelk}
\| |\Psi_{SB}^{\rm bad}(\mathcal{I}_r^{(k)})\rangle\| \le \left(\varepsilon^{(k)}\right)^r~,
\end{equation}
with $\varepsilon^{(k)}$ as in eq.~(\ref{level-k-strength}); the derivation of eq.~(\ref{local-noise-strength-levelk}) is sketched in Appendix A. Furthermore, a level-$k$ simulation in which all $k$-gadgets are good simulates the ideal circuit perfectly. In this sense, repeated level reduction reduces a level-$k$ simulation to an equivalent level-0 simulation while mapping local noise to local noise with a renormalized noise strength $\varepsilon^{(k)}$, and for $\varepsilon < \varepsilon_0$, the  renormalized noise strength becomes negligible for large $k$. The threshold value $\varepsilon_0$ of the noise strength for local noise is of the same order (though not exactly the same) as the threshold for local stochastic noise. We emphasize that, once eq.~(\ref{local-noise-strength}) is established, we can derive eq.~(\ref{local-noise-strength-levelk}) without any further assumptions about the Hamiltonian $H=H_S+H_B+H_{SB}$.

The strength $\varepsilon$ of local noise can be estimated based on the detailed properties of the expansion in eq.~(\ref{expansion-system-bath-Hamiltonian}) of the perturbation $H_{SB}$ in terms of local system operators. For example, in \cite{AGP} the noise was assumed to be ``short range'' in the sense that the perturbation $H_{SB}$ acts collectively on a pair of data qubits only while the ideal system Hamiltonian $H_S$ also couples those two data qubits --- that is, only while the ideal quantum circuit calls for that pair of qubits to undergo a two-qubit gate. For this short-range local noise model, it was shown that eq.~(\ref{local-noise-strength}) is satisfied if we choose
\begin{equation}
\label{short-range}
\varepsilon = \left(\max_{a,t} \| H_{SB}^{(a)}(t)\| \right)\cdot t_0 ~,
\end{equation}
where $t_0$ is the time needed to execute a quantum gate, $\| \cdot \|$ denotes the sup operator norm, and the maximum is over all circuit locations and all times. On the other hand, in \cite{AKP} the noise was assumed to be ``long range'' with $H_{SB}$ coupling each pair of data qubits irrespective of the structure of the ideal circuit. In that case we may write 
\begin{equation}
H_{SB}=\sum_{<ij>} H_{<ij>}~.
\end{equation}
where the sum is over all unordered pairs of system qubits; $H_{<ij>}$ acts collectively on the pair of qubits $<ij>$ and also on the bath. For this long-range local noise model, it was shown that eq.~(\ref{local-noise-strength}) is satisfied if we choose
\begin{equation} 
\label{long-range}
\varepsilon^2 = C\cdot \left( \max_{i,t}\sum_j \| H_{<ij>}\|\right) \cdot t_0 ~,
\end{equation}
where $ C$ is the numerical constant $C = 2e \approx (2.34)^2$. (This is actually a slight improvement over the value of $C$ reported in \cite{AKP}; the improved value can be derived using the reasoning described in Sec.~\ref{subsec:many-marked} below, if we assume $\varepsilon^2 \le e$.)

The origin of eq.~(\ref{short-range}) is easy to understand intuitively \cite{terhal}. If each one of the $r$ specified locations in $\mathcal{I}_r$ is bad, then the perturbation must be inserted at least once in each of these locations, and each insertion reduces the norm of the state by a factor of at least $\|H_{SB}^{(a)}\|$. Inside each location, there is an {\em earliest} insertion of the perturbation that can occur at any time during the duration of the location, a time window of width $t_0$. Integrating over the time of the earliest insertion of the perturbation inside each location, we obtain eq.~(\ref{short-range}).  For the long-range noise model, a single insertion of the perturbation $H_{<ij>}$ can cause two circuit locations to be bad if qubits $i$ and $j$ are participating in separate gates, and therefore the noise strength is correspondingly higher (observe that $\varepsilon^2$ rather than $\varepsilon$ appears on the left-hand-side of eq.~(\ref{long-range})). 

\subsection{Assessment}
\label{subsec:assessment}
The results eq.~(\ref{short-range}) and eq.~(\ref{long-range}) are significant, because they demonstrate that quantum computing is scalable in principle for non-Markovian noise described by a system-bath Hamiltonian. Furthermore, this formulation of the threshold theorem has the noteworthy advantage that the argument works for any bath Hamiltonian $H_B$. The dynamics of the bath does not matter, as long as the perturbation $H_{SB}$ is ``local'' and sufficiently weak. 

However, expressing the threshold condition as in eq.~(\ref{short-range}) or eq.~(\ref{long-range}) has serious drawbacks. First we should note that while in the local stochastic noise model we may interpret the noise strength $\varepsilon$ as an error {\em probability} per gate, in the non-Markovian noise model $\varepsilon$ is really an error {\em amplitude}. Since a probability is a square of an amplitude, requiring $\varepsilon < \varepsilon_0$ in the local noise model is a far more stringent criterion than requiring $\varepsilon < \varepsilon_0$ in the local stochastic noise model. Our analysis yields a much weaker lower bound on the accuracy threshold for the local noise model than for the local stochastic noise model because we pessimistically allow the bad fault paths to add together with a common phase and thus to interfere constructively. Most likely this analysis is far too pessimistic; it is reasonable to expect that distinct fault paths have only weakly correlated phases, and if so, then the modulus of a sum of $N$ fault paths should grow like $\sqrt{N}$ rather than linearly in $N$. That is, if the phases of fault paths can be regarded as random, then we expect the {\em probabilities} of the fault paths, rather than their amplitudes, to accumulate linearly. An important open problem for the theory of quantum fault tolerance is to put this phase-randomization hypothesis on a rigorous footing, and thereby to establish a much higher estimate of the accuracy threshold for local noise. But we will not be addressing this problem in this paper.

There are other drawbacks of the threshold condition eq.~(\ref{short-range}) that we {\em will} try to address, however. One issue is that the norm of the system-bath Hamiltonian is not directly measurable in experiments, and it would be far preferable to state the threshold condition in terms of experimentally accessible quantities, such as the noise power spectrum. In fact, for otherwise reasonable noise models, the norm $\| H_{SB}^{(a)}\|$ could be formally infinite (if for example the system qubits couple to unbounded bath operators such as the quadrature amplitudes of bath oscillators), and in such cases the threshold theorem has little force. 

In more physical terms, an undesirable feature of demanding small $\varepsilon$ where $\varepsilon$ is given by eq.~(\ref{short-range}) is that this condition requires that the very-high-frequency component of the noise be particularly weak, a requirement that seems not to be physically well motivated. To be concrete, suppose that
\begin{equation}
H_{SB}^{(a)}= \mathcal{S}^{(a)}\otimes \mathcal{B}^{(a)}~,
\end{equation}
where $\mathcal{S}^{(a)}$ is a local Hermitian system operator with $\|\mathcal{S}^{(a)}\|=1$ and $\mathcal{B}^{(a)}$ is a Hermitian bath operator. Then combining the condition $\varepsilon < \varepsilon_0$ with eq.~(\ref{short-range}) implies in particular that
\begin{equation}
\label{bath-t-equals-t-prime}
\langle \Psi_{SB}^0|\mathcal{B}^{(a)}(t)\mathcal{B}^{(a)}(t)|\Psi_{SB}^0\rangle= \int_{-\infty}^\infty \frac{d\omega}{2\pi}~ \tilde \Delta(\omega)~ < ~ \varepsilon_0^2/t_0^2~,
\end{equation}
where $\tilde \Delta(\omega)$ is the Fourier transform of the bath's two-point correlation function, defined by
\begin{equation}
\langle\Psi_{SB}^0|\mathcal{B}^{(a)}(t)\mathcal{B}^{(a)}(t')|\Psi_{SB}^0\rangle=\int_{-\infty}^\infty \frac{d\omega}{2\pi} e^{-i\omega(t-t')}\tilde \Delta(\omega)~
\end{equation}
($\mathcal{B}^{(a)}(t)$ denotes the interaction-picture bath operator). Suppose that the fluctuations of the bath variables are Ohmic (and at zero temperature); that is, linear in frequency at low (positive) frequency and exponentially decaying at frequencies large compared to the cutoff frequency $\tau_c^{-1}$:
\begin{equation}
\label{ohmic-spectrum}
\tilde \Delta(\omega)=
\begin{cases} 
2\pi A \omega e^{-\omega\tau_c} & \mbox{if }\omega \ge 0 \\
0 & \mbox{if } \omega < 0
\end{cases}
~,
\end{equation}
where $A$ is a positive dimensionless parameter quantifying the strength of the Ohmic noise. Then the threshold condition implies that 
\begin{equation}
\label{linear-divergence}
\sqrt{A}\cdot \left( \frac{t_0}{\tau_c}\right) < \varepsilon_0~.
\end{equation}
For the case of Ohmic noise, then, the quantity that is required to be small is linearly ``ultraviolet divergent;'' that is, it has a linear sensitivity to the high-frequency cutoff $\tau_c^{-1}$, which may be orders of magnitude higher than the characteristic frequency $t_0^{-1}$ of the ideal computation.  

The extreme sensitivity of the threshold condition to the very-high-frequency noise seems surprising, since one's naive expectation is that noise with zero mean and frequency much larger than $t_0^{-1}$  should nearly average out. This unsatisfying limitation of eq.~(\ref{short-range}), already pointed out in the original paper by Terhal and Burkard \cite{terhal} (and later highlighted by Alicki \cite{alicki} and by Hines and Stamp \cite{stamp}) may just be a shortcoming of the analysis, but conceivably it hints at a deeper problem for quantum fault tolerance. For example, it has been suggested \cite{alicki-horodecki} that during the course of a long quantum computation, an initially benign state of the bath may be pushed toward a far more malicious state that compromises the fault-tolerant protocol. Perhaps high-frequency noise with zero mean, which locally seems incapable of inflicting serious harm, has cumulative global effects that are surprisingly troublesome. Whether or not one suspects that the environment could be so cunning an adversary, stronger rigorous arguments establishing that quantum computing is robust against non-Markovian noise would surely be welcome! 

Our central result in this paper is a new estimate of the noise strength $\varepsilon$ that applies to a Hamiltonian description of Gaussian non-Markovian noise. We will formulate the noise model and state our result in Sec.~\ref{sec:gaussian}, discuss some implications in Sec.~\ref{sec:implications}, and postpone the derivation until Sec.~\ref{sec:derivation}. For this particular important class of noise models, we will be able to state a threshold condition that is less sensitive to very-high-frequency noise, though some sensitivity will still remain. The combinatoric analysis that leads to our result borrows substantially from the derivation in \cite{AKP} of eq.~(\ref{long-range}), though the context is rather different.

\section{Gaussian noise and the threshold condition}
\label{sec:gaussian}

By ``Gaussian noise'' we mean a Hamiltonian noise model where the bath is a set of uncoupled harmonic oscillators, and each system qubit couples to a linear combination of oscillator quadrature amplitudes; hence (in units with $\hbar=1$)
\begin{equation}
H_B= \sum_k \omega_k a_k^\dagger a_k~,
\end{equation}
and
\begin{equation}
\label{gaussian-system-bath}
H_{SB}= \sum_x \sum_\alpha \sigma_\alpha(x) \otimes \tilde\phi_\alpha(x,t)~,
\end{equation}
where $\tilde\phi_\alpha(x,t)$ is the Hermitian operator
\begin{equation}
\label{gaussian-field}
\tilde\phi_\alpha(x,t) = \sum_k \left( g_{k,\alpha}(x,t) a_k + g_{k,\alpha}^*(x,t)a_k^\dagger\right)~.
\end{equation}
Here $x$ is a label indicating a system qubit's position, and $\{\sigma_\alpha(x), \alpha=1,2,3\}$ are the three Pauli operators acting on qubit $x$. The $a_k$'s are annihilation operators for the bath oscillators, satisfying the commutation relation $[a_k,a_{k'}^\dagger]= \delta_{kk'}$, and $g_{k,\alpha}(x,t)$ is a complex coupling parameter that determines how strongly oscillator $k$ couples to qubit $x$ at time $t$. 

As explained in Sec.~\ref{subsec:local-noise} (see also Sec.~\ref{subsec:initial-state}), we may assume without loss of generality that the bath is prepared in a pure state $|\Psi_B^0\rangle$ at the beginning of the computation. The Hamiltonian $H_B +H_{SB}$, along with the choice of the bath's initial state $|\Psi_B^0\rangle$, defines our noise model. The bath fluctuations will be Gaussian if the state $|\Psi_B^0\rangle$ is a Gaussian state (that is, a generalized ``squeezed'' state) of the oscillator bath --- a purified thermal state is a special case of such a squeezed state.

It is useful to define the ``interaction picture'' bath operator $\phi_\alpha(x,t)$ as
\begin{equation}
\phi_\alpha(x,t)= e^{i H_{B}t} \tilde\phi_\alpha(x,t)e^{-iH_{B} t} = 
\sum_k \left( g_{k,\alpha}(x,t) a_ke^{-i\omega_k t} + g_{k,\alpha}^*(x,t)a_k^\dagger e^{i\omega_k t}\right)~,
\end{equation}
and to define the bath's two-point correlation function as
\begin{equation}
\Delta(\alpha_1,x_1,t_1;\alpha_2,x_2,t_2)= \langle \Psi_B^0|\phi_{\alpha_1}(x_1,t_1)\phi_{\alpha_2}(x_2,t_2)|\Psi_B^0\rangle~.
\end{equation}
We will sometimes use the abbreviated notation $\phi(1)$ for $\phi_{\alpha_1}(x_1,t_1)$ and $\Delta(1,2)$ for $\Delta(\alpha_1,x_1,t_1;\alpha_2,x_2,t_2)$; we also define
\begin{equation}
\label{delta-bar}
|\bar\Delta(1,2)|= \sum_{\alpha_1,\alpha_2} |\Delta(\alpha_1,x_1,t_1;\alpha_2,x_2,t_2)|~.
\end{equation}

\begin{figure*}[t]
\begin{center}
\vspace{-.5cm}
\includegraphics[width=16cm,keepaspectratio]{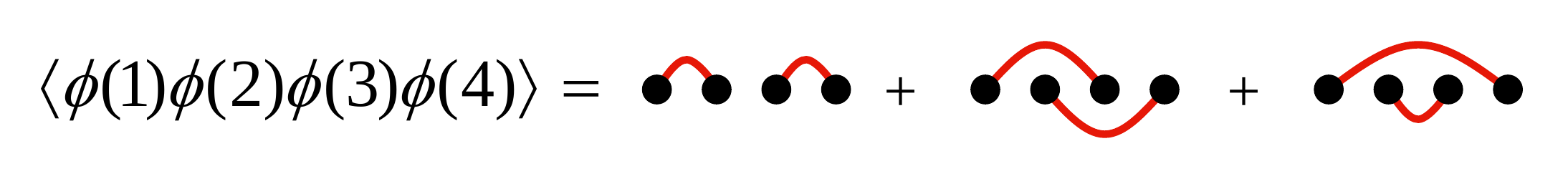}
\vspace{-.5cm} 
\end{center}
\caption{\label{fig:four-point} The four-point correlation function for a free field can be expressed in terms of products of two-point correlation functions by summing over all ``contractions,'' where each contraction divides the four points into two unordered pairs.
         }
\end{figure*}

When we say that the noise is Gaussian, we mean that the bath variable $\phi_\alpha(x,t)$ obeys Gaussian statistics:  all $n$-point bath correlation functions vanish for $n$ odd, and the $2n$-point function can be expressed in terms of two-point functions. Using $\langle ~\cdot ~\rangle$ to denote the expectation value in the state $|\Psi_B^0\rangle$, Gaussian statistics implies that
\begin{equation}
\langle \phi(1)\phi(2)\phi(3) \cdots \phi(2n)\rangle = \sum_{\rm contractions} \Delta(i_1,i_2)\Delta(i_3,i_4)\cdots  \Delta(i_{2n-1},i_{2n})~,
\end{equation}
where summing over ``contractions'' means summing over the $(2n)!/2^nn!$ ways to divide the labels $1,2,3, \dots 2n$ into $n$ unordered pairs. For example, if $\phi$ is a Gaussian variable, then the four-point function is 
\begin{equation}
\langle \phi(1)\phi(2)\phi(3)\phi(4)\rangle=\Delta(1,2)\Delta(3,4)+\Delta(1,3)\Delta(2,4)+\Delta(1,4)\Delta(2,3)~,
\end{equation}
as illustrated in Fig.~\ref{fig:four-point}.
This expansion of the $2n$-point function in terms of two-point functions is sometimes called ``Wick's theorem.''

Now we can state our main result: Gaussian noise obeys the local noise condition eq.~(\ref{local-noise-strength}), with noise strength
\begin{equation}
\label{result}
\varepsilon^2= C \cdot \max_{\rm Loc} \left(\int_{1, {\rm Loc}}\int_{2, {\rm All}}|\bar\Delta(1,2)|\right)~,
\end{equation}
where $C=2e\approx(2.34)^2$ is the numerical constant defined earlier (and where we have assumed $\varepsilon^2 \le e$). Here $\int_{1,\rm {Loc}}$ indicates that one leg $(x_1,t_1)$ of the two-point function is integrated over a single location in the circuit: $x_1$ is summed over the qubits participating in a particular gate, and $t_1$ is integrated over the time interval in which that gate is executed. And $\int_{2,\rm{All}}$ indicates that the other leg $(x_2,t_2)$ of the two-point function is summed over all system qubits and integrated over the entire duration of the computation. The maximum is with respect to all possible circuit locations for $(x_1,t_1)$. The threshold condition $\varepsilon < \varepsilon_0$, with $\varepsilon$ given by eq.~(\ref{result}), now becomes a condition on the two-point correlation function of the bath. We note that the ordering of the operators $\phi(1)$ and $\phi(2)$ does not matter in eq.~(\ref{result}) because $|\Delta(1,2)|=|\Delta(2,1)|$; changing the ordering modifies only the phase of $\Delta(1,2)$, not its modulus. 

Another noteworthy feature is that our estimate of $\varepsilon$ applies for an arbitrary system Hamiltonian. This property may seem unexpected at first, as we know that in some settings the damage caused by the noise can depend on the relation between the energy spectrum of the $H_S$ and the power spectrum of the noise. For example, the spontaneous decay rate for a qubit with energy splitting $\hbar\omega$ depends on the noise power at circular frequency $\omega$. How, then, can our threshold condition depend only on the noise spectrum and not on the energy spectrum of $H_S$? The answer is that by taking the modulus $|\bar\Delta(1,2)|$ of the bath two-point function in eq.~(\ref{result}) we are already being maximally pessimistic about how the spectrum of $H_S$ matches the noise power spectrum. Thus there are both advantages and disadvantages in formulating a threshold condition that is general enough to apply for any ideal system Hamiltonian. On the one hand we find a criterion for scalable quantum computing that can be stated easily and rigorously proved by a reasonably simple argument. On the other hand, the price of such rigor is that our stated criterion may be far more demanding than it really needs to be. 

The crucial assumption in the derivation of eq.~(\ref{result}) is eq.~(\ref{gaussian-system-bath}), where $\tilde\phi_\alpha(x,t)$ is a ``free field,'' {\em i.e.}, obeys Gaussian statistics; thus eq.~(\ref{gaussian-field}) could be regarded as merely a general phenomenological representation of a Gaussian field, and not necessarily as a fundamentally accurate microscopic description of the bath. Caldeira and Leggett \cite{leggett} have argued that noise is expected to be Gaussian, at least to an excellent approximation, in a wide variety of realistic physical settings where the system is weakly coupled to many environmental degrees of freedom.

If the initial state of the bath is a thermal state with inverse temperature $\beta=1/kT$, then the mean occupation number of each oscillator is determined by the Bose-Einstein distribution function; we have
\begin{eqnarray}
&&\langle \Psi_B^0|a_k^\dagger a_{k'}|\Psi_B^0\rangle= \frac{\delta_{kk'}}{e^{\beta\omega_k}-1}
= \langle\Psi_B^0|a_k a_{k'}^\dagger|\Psi_B^0\rangle - 1~,\nonumber\\
&&\langle \Psi_B^0|a_k a_{k'}|\Psi_B^0\rangle= 0 = \langle \Psi_B^0|a_k^\dagger a_{k'}^\dagger|\Psi_B^0\rangle~,
\end{eqnarray}
and therefore
\begin{eqnarray}
\Delta(\alpha_1,x_1,t_1;\alpha_2,x_2,t_2)= 
&&\frac{1}{2}\sum_k g_{k,\alpha_1}(x_1,t_1)g_{k,\alpha_2}^*(x_2,t_2)e^{-i\omega_k(t_1-t_2)}\left(\coth(\beta\omega_k/2)+1\right)\nonumber\\
+ &&\frac{1}{2}\sum_k g_{k,\alpha_1}^*(x_1,t_1)g_{k,\alpha_2}(x_2,t_2)e^{i\omega_k(t_1-t_2)}\left(\coth(\beta\omega_k/2)-1\right)~.
\end{eqnarray}
Just to be concrete, consider the case where the noise is stationary and spatially uncorrelated --- each qubit has a time-independent coupling to its own independent oscillator bath (though admittedly these are dubious assumptions when multi-qubit gates are executed). Then
\begin{equation}
\Delta(\alpha_1,x_1,t_1;\alpha_2,x_2,t_2)= \delta_{x_1 x_2}\Delta(\alpha_1,x_1,t_1;\alpha_2,x_1,t_2)~,
\end{equation}
where
\begin{equation}
\Delta(\alpha_1,x_1,t_1;\alpha_2,x_1,t_2)= \int_{-\infty}^\infty\frac{d\omega}{2\pi}e^{-i\omega (t_1-t_2)}\tilde \Delta_{\alpha_1\alpha_2}(x_1,\omega)
\end{equation}
and 
\begin{equation}
\tilde \Delta_{\alpha_1\alpha_2}(x_1,\omega)=
\begin{cases} 
\pi J_{\alpha_1,\alpha_2}(x_1,\omega)\left(\coth(\beta\omega/2) +1\right)& \mbox{if }\omega > 0 \\
\pi J_{\alpha_1,\alpha_2}^*(x_1,\omega)\left(\coth(\beta\omega/2) -1\right) & \mbox{if } \omega < 0
\end{cases}
~.
\end{equation}
Here $J_{\alpha_1,\alpha_2}(x_1,\omega)$ is the Hermitian matrix 
\begin{equation}
J_{\alpha_1,\alpha_2}(x_1,\omega)= \sum_k \delta(\omega-\omega_k)g_{k,\alpha_1}(x_1) g_{k,\alpha_2}^*(x_1)~.
\end{equation}
The function $J_{\alpha_1,\alpha_2}(x_1,\omega)$ is the spin-polarization-dependent power spectrum of the noise acting on qubit $x_1$. If the energy splitting $\hbar\omega$ of the qubit is tunable, this function can be measured by observing the qubit's relaxation rate as a function of the energy splitting and the polarization. In principle, multi-qubit correlations in the noise can also be measured using quantum process tomography. 

\section{Some implications}
\label{sec:implications}
Before presenting our derivation of eq.~(\ref{result}) in Sec.~\ref{sec:derivation}, we will discuss a few of its implications. 

\subsection{Dimensional criterion}
\label{sec:dimension}
Our expression for $\varepsilon$ in eq.~(\ref{result}) involves a formal integration over all space and time. If the bath correlations decay slowly in space or time, this integral might diverge in the limit of a computation that is very wide, very deep, or both. In that case, our ``threshold condition'' cannot be satisfied asymptotically, and we cannot conclude that quantum computation is scalable. On the other hand, if the integral converges ``in the infrared,'' then the threshold condition has value, as it establishes scalability if the coupling of the system to the bath is sufficiently weak. As long as $\varepsilon$ is finite, we can make it as small as we please by weakening the coupling of the qubits to the bath, {\em i.e.}, by rescaling the perturbation $H_{SB}$, or equivalently by rescaling the field $\phi_\alpha(x,t)$. 

What is the criterion for infrared convergence? Let us suppose that the qubits are uniformly distributed in $D$-dimensional space, and that the bath fluctuations are ``critical;'' {\em i.e.}, algebraically decaying in space and in time. We say that the scale dimension of the field $\phi$ is $\delta$ and the dynamical critical exponent is $z$ if, for large scale factor $\lambda$, the bath two-point function scales according to
\begin{equation}
\Delta(\lambda x_1, \lambda^z t_1; \lambda x_2, \lambda^z t_2)\sim 
\lambda^{-2\delta} \Delta(x_1,t_1;x_2,t_2)~;
\end{equation}
thus the time $t$ scales like $z$ powers of the spatial distance $x$. This means that the integral of the two-point function scales as
\begin{equation}
\int dt~ d^Dx~ |\Delta(x,t;0,0)| \sim R^{D+z-2\delta}~,
\end{equation}
where $R$ is an infrared cutoff. Convergence in the infrared (finiteness of the limit $R\to\infty$) is ensured provided that
\begin{equation}
D+z < 2\delta~;
\end{equation}
if this criterion is satisfied, then scalable quantum computing is achievable at weak coupling. If it is not satisfied, then scalable quantum computing might still be possible, but our version of the threshold theorem does not guarantee it. The same criterion was previously stated by Novais {\em et al.} \cite{novais1,novais2}, though without rigorous justification.

\subsection{Almost-Markovian noise}
The noise is Markovian if the bath immediately ``forgets'' any quantum information it receives, so that the information never returns to the system. Though this is never strictly the case, it can be true to an excellent approximation if the characteristic correlation time of the bath is very short compared to the time resolution with which we monitor the system's behavior. In the Gaussian noise model, the noise is Markovian if the bath's two-point correlation function is proportional to a delta function of the time difference, 
\begin{equation}
\Delta(t_1,x_1;t_2,x_2)\propto \delta(t_1 - t_2)~.
\end{equation}
We could say that the noise is ``almost Markovian'' if the correlation function $\Delta$ is a sharply peaked function of the time difference, {\em e.g.}, with width $\tau_c$ much less than the duration $t_0$ of a single quantum gate. In that case, our expression for the noise strength becomes
\begin{equation}
\label{almost-markovian}
\varepsilon^2= C\cdot \max_{\rm Loc} \left(\int_{1, {\rm Loc}}\int_{2, {\rm All}}|\bar\Delta(1,2)|\right)\approx \Gamma t_0~;
\end{equation}
for each fixed value of $t_1$, the sharply peaked $t_2$ integral generates the factor $\Gamma$, and then integrating $t_1$ over the duration of the location generates the factor $t_0$. 

We may interpret $\Gamma$ as an error rate per unit time, and $\Gamma t_0$ as an error probability per gate. But note that the noise strength $\varepsilon$ is not this error rate, but rather its square root $\left(\Gamma t_0\right)^{1/2}$, in effect the {\em amplitude} of the error. In the Markovian case, fault paths really {\em do} decohere, and errors {\em can} be assigned probabilities rather than amplitudes. But our derivation of eq.~(\ref{result}) is too general and insufficiently clever to exploit this property; hence our threshold condition requires the error amplitude rather than its square to be less than $\varepsilon_0$. 

Despite this deficiency, at least our threshold criterion for local noise, when applied to the almost-Markovian case, improves on the operator norm criterion eq.~(\ref{short-range}). If the two-point function has a narrow peak of width $\tau_c$ whose integral is $\Gamma$, then the {\em height} of the peak is of order $\Gamma/\tau_c$, and this peak height can be interpreted as the norm squared of the system-bath Hamiltonian, as in eq.~(\ref{bath-t-equals-t-prime}). Thus eq.~(\ref{short-range}) estimates the noise strength as
\begin{equation}
\label{almost-short-range}
\varepsilon \sim \sqrt{\Gamma/\tau_c}\cdot t_0~, 
\end{equation}
which is even more pessimistic than eq.~(\ref{almost-markovian}). The estimate eq.~(\ref{almost-short-range}) diverges as the ultraviolet cutoff $\tau_c^{-1}$ is removed. But the estimate eq.~(\ref{almost-markovian}) depends on the area under the peak rather than its height, and so has a smooth limit as $\tau_c\to 0$.

\subsection{Ohmic noise}
To further explore the sensitivity to high-frequency noise of our estimated noise strength, let us consider the Ohmic case, as in Sec.~\ref{subsec:assessment}. If the Fourier transform $\tilde\Delta(\omega)$ of the two-point correlation function is given by eq.~(\ref{ohmic-spectrum}), then the real-time correlation function is
\begin{equation}
\Delta(t)=\int_{-\infty}^\infty \frac{d\omega}{2\pi} e^{-i\omega t}\tilde \Delta(\omega)
= \frac{-A}{(t-i\tau_c)^2}~.
\end{equation} 
The function $\Delta(t)$  has a short-time singularity at $t=0$ that is regulated by the cutoff $\tau_c$, but 
the real and imaginary parts of $\Delta(t)$ both oscillate, so that its time integral  vanishes:
\begin{equation}
\int_{-\infty}^\infty dt~\Delta(t) = \tilde \Delta(\omega=0)=0~.
\end{equation}

However the estimated noise strength $\varepsilon$, which is required to be small by the threshold condition, depends on the integral of the {\em modulus} of $\Delta(t)$, 
\begin{equation}|\Delta(t)|= \frac{A}{t^2 + \tau_c^2}~,
\end{equation}
which is of course nonnegative and has a nonvanishing time integral; the estimated noise strength is
\begin{equation}
\label{ohmic-integral}
\varepsilon = \left(C\cdot \int_{\rm Loc} dt \int_{\rm All}  ds ~|\Delta(t-s)|\right)^{1/2}=\sqrt{\pi C A}\cdot \left(\frac{t_0}{\tau_c}\right)^{1/2}~.
\end{equation} 
This estimate is ultraviolet divergent, but comparing to eq.~(\ref{linear-divergence}) we see that the divergence has been improved from linear to square-root dependence on the ultraviolet cutoff $\tau_c^{-1}$.

Despite the improvement, the surviving ultraviolet sensitivity in this estimate of $\varepsilon$ (for the case of Ohmic noise) is troubling, as it significantly reduces the class of noise models for which we can conclude that quantum computing is scalable. Therefore it is important to understand the origin of the ultraviolet divergence. One might suspect at first that the ultraviolet sensitivity arises because the range of the $dt$ integral in eq.~(\ref{ohmic-integral}) is a window of width $t_0$ with a sharp boundary. But in fact, for Ohmic noise the sharp boundary generates only a mild logarithmic ultraviolet divergence, not a power divergence. The actual reason for the power divergence is that we have pled complete ignorance regarding the frequency spectrum of the ideal system Hamiltonian $H_S$. Therefore, we are required to be maximally pessimistic about how the oscillating phase of the wave function arising from the ideal system dynamics matches with the phase of the bath fluctuations. For that reason our estimate of $\varepsilon$ involves an integral of the modulus of $\Delta(t)$ rather than $\Delta(t)$ itself.

With further assumptions about the ideal system dynamics we ought to be able to exclude this highly pessimistic scenario, leading to an estimate of $\varepsilon$ with milder ultraviolet sensitivity. A natural idea is to attempt a ``renormalization group improvement'' of the noise model; that is, to ``coarse grain'' in time, stretching the short-time cutoff $\tau_c$, while adjusting the bath fluctuations to keep invariant the effect of the noise on the system. Formally Ohmic noise is ``marginal,'' meaning that the naive renormalization-group scaling generates only logarithmic cutoff dependence, not the square-root dependence found in eq.~(\ref{ohmic-integral}). However, rigorously justifying this naive scaling turns out to be technically difficult, in part because $H_{SB}$ couples the system operators to {\em unbounded} bath operators in the Gaussian noise model. It might be interesting to see if further technical assumptions (which one would hope to justify {\em a posteriori}) about the system-bath state $|\Psi_{SB}(t)\rangle$ during the course of the computation would lead to a less ultraviolet-sensitive threshold condition, but we have not yet succeeded in finding useful results with this character. 

\section{Derivation}
\label{sec:derivation}
In this section we will derive eq.~(\ref{result}). Our task is to estimate a value of $\varepsilon$ such that
\begin{equation}
\||\Psi_{SB}^{\rm bad}(\mathcal{I}_r)\rangle\|^2 = 
\langle \Psi_{SB}^{\rm bad}(\mathcal{I}_r)|\Psi_{SB}^{\rm bad}(\mathcal{I}_r)\rangle
= \langle \Psi_{SB}^0 |U_{SB}^{\rm bad}(\mathcal{I}_r)^\dagger U_{SB}^{\rm bad}(\mathcal{I}_r)|\Psi_{SB}^0\rangle \le \varepsilon^{2r}~
\end{equation}
(see eq.~(\ref{local-noise-strength})). Here $U_{SB}$ is the joint system-bath time evolution operator from the beginning of the computation until just before the measurements that will read out the final result, and $U_{SB}^{\rm bad}(\mathcal{I}_r)$ denotes the sum of all the terms in the perturbation expansion of $U_{SB}$ such that the perturbation $H_{SB}$ is inserted at least once in each of the $r$ specified locations in the set $\mathcal{I}_r$. The initial state of the system and bath is assumed to be the pure product state $|\Psi_{SB}^0\rangle = |\Psi_{S}^0\rangle\otimes |\Psi_B^0\rangle$, where the bath's state $|\Psi_B^0\rangle$ is Gaussian; that is, the expectation values of the bath operators $\{\phi_\alpha(x,t)\}$ obey Gaussian statistics in this state. For now we assume that system qubits are prepared only at the start of the computation and measured only at the end, with evolution governed by the Hamiltonian $H=H_S+H_B+H_{SB}$ in between; this assumption can be relaxed, as we will discuss in Sec.~\ref{sec:generalizations}. 

\begin{figure*}[t]
\begin{center}
\vspace{-.5cm}
\includegraphics[height=8cm,keepaspectratio]{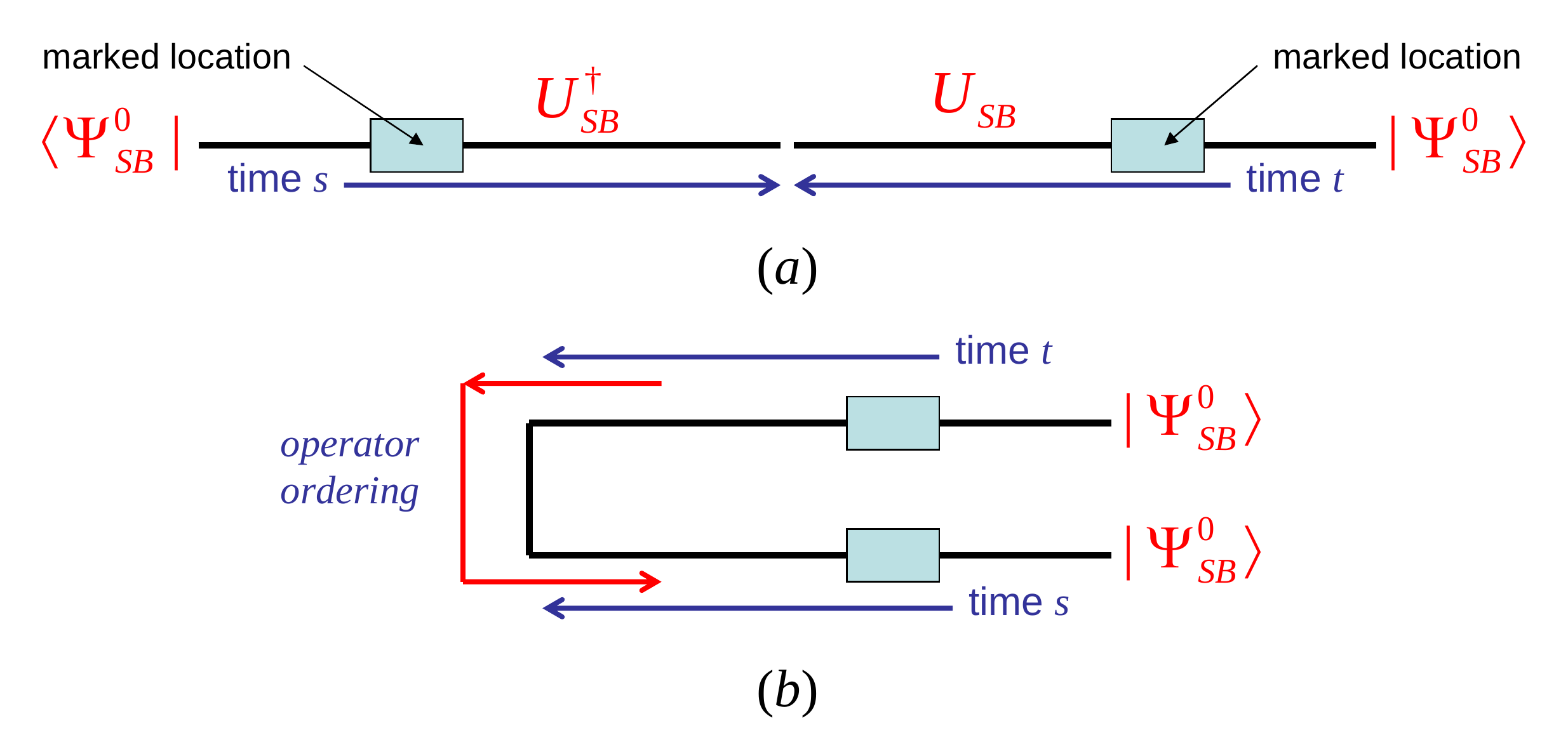}
\vspace{-.7cm} 
\end{center}
\caption{\label{fig:keldysh} The diagram in ($a$) represents the norm squared of the system-bath state $U_{SB}|\Psi_{SB}^0\rangle$. By bending it into a hairpin shape we obtain the ``Keldysh diagram'' shown in ($b$). A ``marked location'' in the circuit appears twice in the Keldysh diagram, once on the upper branch (as a contribution to $U_{SB}$), and once on the lower branch (as a contribution to $U_{SB}^\dagger$).
         }
\end{figure*}

\subsection{Keldysh diagrams}
The terms in the perturbation expansion can be associated with diagrams, where in each diagram the perturbation $H_{SB}$ is inserted at a specified set of points in spacetime. We may think of the sum of these diagrams as representing the expectation in the state $|\Psi_{SB}^0\rangle$ of the product of the forward evolution operator of the system and bath ({\em i.e.}, $U_{SB}$), from the initial to the final time, followed by the backward evolution operator ({\em i.e.}, $U_{SB}^\dagger$), from the final to the initial time. It is convenient to fold the diagram into a hairpin shape as in Fig.~\ref{fig:keldysh}, so that the diagram has two branches that are aligned in time. The upper branch represents the evolution forward in time; here the inserted perturbations are ``time-ordered,'' meaning that operators inserted at later times act after operators inserted at earlier times. The lower branch represents the evolution backward in time; here the inserted perturbations are ``anti-time-ordered,'' meaning that operators inserted at earlier times act after operators inserted at later times. Furthermore, all operators inserted on the lower branch act after operators inserted on the upper branch. Diagrams with this structure are sometimes called ``Keldysh diagrams.''

In each diagram, the evolution of system and bath is governed by the uncoupled Hamiltonian $H_0=H_S+H_B$ in between successive insertions of $H_{SB}$. Since $|\Psi_{SB}^0\rangle$ is a product state, the diagram's contribution to the expectation value factorizes into the product of a system expectation value and a bath expectation value. Consider a diagram where the operator $\sigma_{\alpha_j}\otimes \tilde\phi_{\alpha_j}$ is inserted on the upper branch acting on qubit $x_j$ at time $t_j$, for $j=1,2,3,\dots, n$, and the operator $\sigma_{\beta_k}\otimes \tilde\phi_{\beta_k} $ is inserted on the lower branch acting on qubit $y_k$ at time $s_k$, for $k=1,2,3,\dots, m$. Taking into account the uncoupled evolution in between insertions, and the Keldysh operator ordering rules (where $t_n > t_{n-1} > \cdots > t_1$ and $s_m < s_{m-1} < \cdots < s_1$), this diagram's contribution to the expectation value $\langle \Psi_{SB}^0|U_{SB}^\dagger U_{SB}|\Psi_{SB}^0\rangle$ is
\begin{eqnarray}
i^m(-i)^n\times &&\langle \Psi_S^0| \sigma_{\beta_m}(y_m,s_m) \cdots \sigma_{\beta_1}(y_1,s_1)\sigma_{\alpha_n}(x_n,t_n)\cdots \sigma_{\alpha_1}(x_1,t_1)|\Psi_S^0\rangle\nonumber\\
\times && \langle\Psi_B^0| \phi_{\beta_m}(y_m,s_m) \cdots \phi_{\beta_1}(y_1,s_1)\phi_{\alpha_n}(x_n,t_n)\cdots \phi_{\alpha_1}(x_1,t_1) |\Psi_B^0\rangle~.
\end{eqnarray}
Here $\sigma_\alpha(x,t)= U_S(t)^\dagger\sigma_\alpha(x)U_S(t)$ and $\phi_\alpha(x,t)=U_B(t)^\dagger\tilde\phi_\alpha(x,t)U_B(t)$ are the ``interaction picture'' operators that evolve according to the uncoupled system-bath dynamics. Using the Gaussian statistics ({\em i.e.}, ``Wick's theorem''), the bath expectation value can be expressed as a sum of products of Keldysh-ordered two-point correlation functions. Summing $\{\alpha_1,\alpha_2,\dots \alpha_n\}$ and $\{\beta_1,\beta_2,\dots,\beta_m\}$ from $1$ to $3$, summing $\{x_1,x_2,\dots x_n\}$ and $\{y_1, y_2, \dots y_m\}$ over all qubits, integrating $\{t_1,t_2,\dots\}$ and $\{s_1,s_2,\dots, s_m\}$ over the interval from the initial to the final time, and finally summing $n$ and $m$ from $0$ to $\infty$, we would recover the full expectation value $\langle \Psi_{SB}^0|U_{SB}^\dagger U_{SB}|\Psi_{SB}^0\rangle=1$. More precisely, to generate the full system-bath evolution operator $U_{SB}$, for each fixed $n$ we sum $\{(x_1,t_1),(x_2,t_2),(x_3,t_3), \dots, (x_n,t_n)\}$ over all time-ordered sets of $n$ spacetime positions inside the circuit. This is equivalent to integrating each $(x_j,t_j)$ over all spacetime, and then dividing by $n!$ to compensate for the overcounting of the sets (each set has been included $n!$ times). Similarly, to generate $U_{SB}^\dagger$, for each fixed $m$ we sum $\{(y_1,s_1),(y_2,s_2),(y_3,s_3), \dots, (y_m,s_m)\}$ over all anti-time-ordered sets of $m$ spacetime positions inside the circuit. 

\subsection{One marked location}
\begin{figure*}[t]
\begin{center}
\vspace{-.5cm}
\includegraphics[width=9.5cm,keepaspectratio]{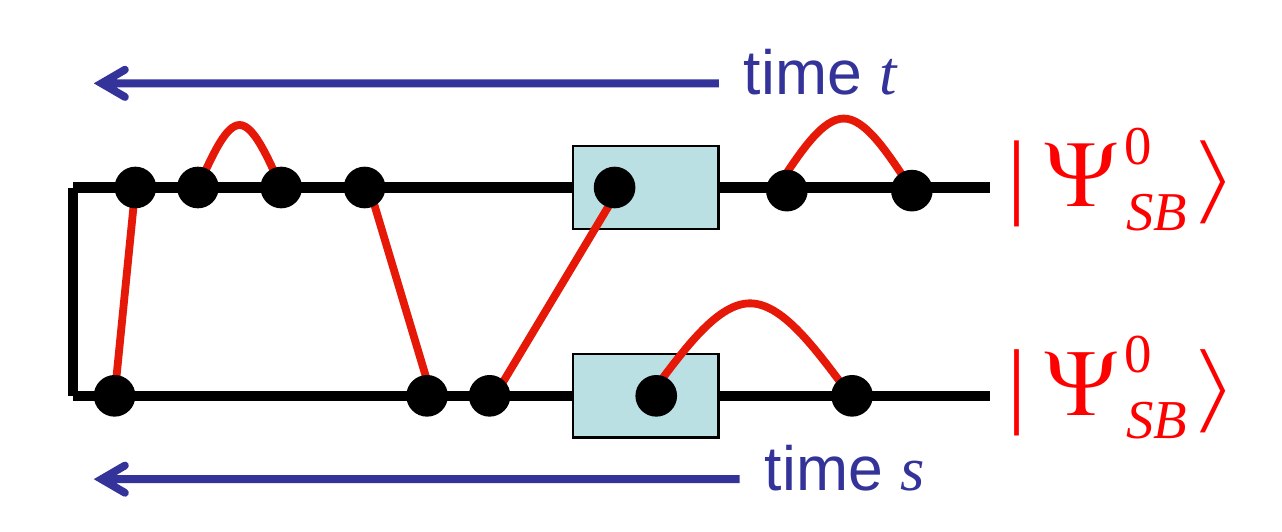}
\vspace{-.7cm} 
\end{center}
\caption{\label{fig:r-equals-one} A Keldysh diagram contributing to $\langle \Psi_{SB}^{\rm bad}(\mathcal{I}_{r=1})|\Psi_{SB}^{\rm bad}(\mathcal{I}_{r=1})\rangle$, where black dots are insertions, red lines are contractions, and the shaded rectangle on each branch indicates the one marked location. In this diagram, the marked location is ``bad'' because there is an insertion inside the marked location on each branch.
         }
\end{figure*}

But we do not want to sum all the diagrams; instead we want to sum all and only those such that all of the $r$ locations in the set $\mathcal{I}_r$ are bad on both the upper and lower branches. Let us first consider the case $r=1$, where one particular circuit location has been ``marked'' as bad. To get a useful bound, it is helpful to organize this sum in a particular way. Because the marked location is bad, there must be at least one insertion of the perturbation inside this location on both the upper and lower branch, as in Fig.~\ref{fig:r-equals-one}. Therefore, there must be an {\em earliest} insertion inside the marked location on each branch. Also, if the marked location is a two-qubit gate, then the earliest insertion could act on either one of the two qubits. For now, let us fix on each branch the time of the earliest insertion inside the marked location, the qubit on which the earliest insertion acts, and the corresponding Pauli operator. Later on we will integrate the time of the earliest insertion over the marked location, and also sum over Pauli operators and the qubits at the location, but not yet.

\begin{figure*}[t]
\begin{center}
\includegraphics[height=5cm,keepaspectratio]{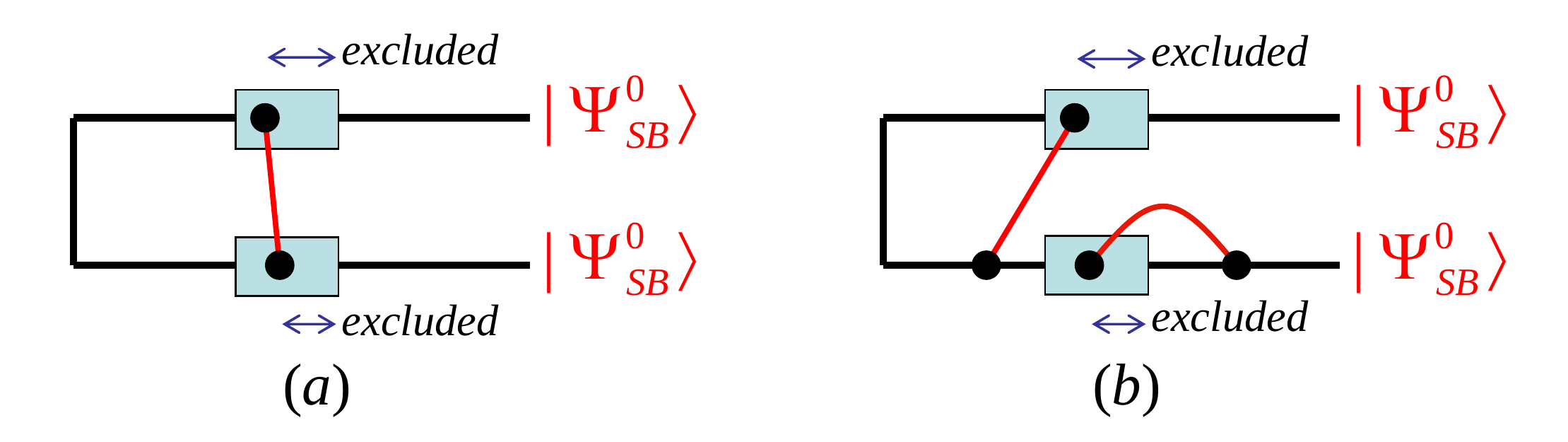}
\vspace{-.7cm} 
\end{center}
\caption{\label{fig:class-one-and-two} Skeleton Keldysh diagrams contributing to $\langle \Psi_{SB}^{\rm bad}(\mathcal{I}_{r=1})|\Psi_{SB}^{\rm bad}(\mathcal{I}_{r=1})\rangle$, showing the earliest insertion of $H_{SB}$ inside the marked location on both branches. For the class 1 diagram shown in ($a$), the earliest insertions on the two branches are contracted with one another, and for the class 2 digram shown in ($b$), the earliest insertions are contracted with other insertions elsewhere. Other diagrams in each class are obtained by dressing the skeleton diagrams with additional insertions and contractions, except that no insertions are allowed inside the marked locations at times before the earliest insertion.
         }
\end{figure*}

With the earliest insertion fixed on each branch, and after expanding the bath expectation value in terms of bath two-point functions, we can identify two classes of diagrams. In ``class 1'' diagrams, the earliest insertions on the two branches are contracted with one another, and in ``class 2'' diagrams they are not, as shown in Fig.~\ref{fig:class-one-and-two}. We will find upper bounds on the sum of all the diagrams in each class.

\subsubsection{Class 1 diagrams}
\label{subsubsec:class1}
First consider the class 1 diagrams. Each diagram in the class has as a factor the two-point function $\Delta(\beta,y,s;\alpha,x,t)$, where $\sigma_\alpha$ acts on qubit $x$ at time $t$ on the upper branch, and $\sigma_\beta$ acts on qubit $y$ at time $s$ on the lower branch. The simplest diagram in the class, which we will call the ``skeleton'' digram, has only two insertions and one contraction; its value is 
\begin{eqnarray}
\label{skeleton}
\langle \Psi_S^0| \sigma_{\beta}(y,s) \sigma_{\alpha}(x,t)|\Psi_S^0\rangle
\times \langle\Psi_B^0| \phi_{\beta}(y,s) \phi_{\alpha}(x,t) |\Psi_B^0\rangle~.
\end{eqnarray}
The other class 1 diagrams are obtained by ``dressing'' this skeleton in all possible ways, by adding further insertions and contractions. 

However, remember that we have fixed $t$ and $s$ to be the times of the earliest insertions of the perturbation on the upper and lower branches respectively. Therefore, an additional insertion is ``legal'' only if it avoids times earlier than $t$ inside the marked location on the upper branch, and times earlier than $s$ inside the marked location on the lower branch. With this proviso, all class 1 diagrams arise when we dress the skeleton class 1 diagram with all possible additional legal insertions and contractions.

The class 1 diagrams can be summed up and expressed in a compact form. For this purpose we introduce what we call the ``hybrid picture,'' which is in a sense intermediate between the interaction and Heisenberg pictures. Let us define the ``hybrid Hamiltonian'' $H^{\rm hyb}$ by
\begin{equation}
H^{\rm hyb}=
\begin{cases} 
H_S + H_B & \mbox{in each marked location prior to the earliest insertion, }\\
H_S+H_B +H_{SB} & \mbox{elsewhere. }
\end{cases}
\end{equation}
That is, in the hybrid Hamiltonian, the perturbation $H_{SB}$ ``turns off'' inside the marked location before time $t$ on the upper branch and before time $s$ on the lower branch. When we sum up all the legal insertions and contractions, the value eq.~(\ref{skeleton}) of the skeleton diagram is transformed into
\begin{eqnarray}
\label{class1}
\langle \Psi_{SB}^0| \sigma_{\beta}^{\rm hyb}(y,s) \sigma_{\alpha}^{\rm hyb}(x,t)|\Psi_{SB}^0\rangle
\times \langle\Psi_B^0| \phi_{\beta}(y,s) \phi_{\alpha}(x,t) |\Psi_B^0\rangle~.
\end{eqnarray}
Here the interaction picture operator $\sigma_\alpha(x,t)$ has been replaced by the hybrid-picture operator $\sigma_{\alpha}^{\rm hyb}(x,t)=U_{SB}^{\rm hyb}(t)^\dagger \sigma_\alpha(x)U_{SB}^{\rm hyb}(t)$, and furthermore the expectation value of the system operator is now evaluated in the system-bath state $|\Psi_{SB}^0\rangle$ rather than the system state $|\Psi_{S}^0\rangle$. If the expression eq.~(\ref{class1}) is expanded in powers of $H^{\rm hyb}_{SB}$, all the legal insertions and only the legal insertions are generated. And for each choice of insertions, evaluating the bath expectation value using Wick's theorem yields a sum over all the contractions included in class 1. Thus, the assumption that the bath fluctuations are Gaussian is crucial for the derivation of eq.~(\ref{class1}).

Therefore, we obtain an exact expression for the sum of all the diagrams in class 1 from eq.~(\ref{class1}) by now integrating the earliest insertions on both branches over the marked location, finding
\begin{eqnarray}
\label{class1-sum}
&&\sum \mbox{Class 1 Diagrams}\nonumber\\
&&=\sum_{x,y \in {\rm Loc}} \int_{\rm Loc}dt \int_{\rm Loc} ds \sum_{\alpha,\beta}
\langle \Psi_{SB}^0| \sigma_{\beta}^{\rm hyb}(y,s) \sigma_{\alpha}^{\rm hyb}(x,t)|\Psi_{SB}^0\rangle
\langle\Psi_B^0| \phi_{\beta}(y,s) \phi_{\alpha}(x,t) |\Psi_B^0\rangle~.
\end{eqnarray}
Now, the operator $\sigma_\alpha^{\rm hyb}(x,t)$ differs from the Pauli operator $\sigma_\alpha(x)$ by a mere unitary change of basis, and therefore has sup norm $\|\sigma_\alpha^{\rm hyb}(x,t)\|=1$. From eq.~(\ref{class1-sum}) we then conclude that
\begin{eqnarray}
\label{class1-sum-bound}
&& \left|\sum \mbox{Class 1 Diagrams}\right|\nonumber\\
&& \le \sum_{x,y \in {\rm Loc}} \int_{\rm Loc}dt \int_{\rm Loc} ds \sum_{\alpha,\beta}
\left|\langle\Psi_B^0| \phi_{\beta}(y,s) \phi_{\alpha}(x,t) |\Psi_B^0\rangle\right|~
= \int_{1,{\rm Loc}}\int_{2,{\rm Loc}}|\bar \Delta(1,2)|~,
\end{eqnarray}
in the notation of eq.~(\ref{delta-bar}). This is our bound on the sum of all class 1 diagrams.

Note that in eq.~(\ref{class1-sum}) the integrand is the product of a bath two-point correlation function and a ``hybridized'' system two-point correlation function. If the bath correlation function has a high-frequency component and the system correlation function does not, then the contribution to the time integral arising from the high-frequency bath fluctuations may be strongly suppressed. But the estimate in eq.~(\ref{class1-sum-bound}) is very crude --- it applies irrespective of the frequency spectrum of the system correlation function --- and we could get a better estimate if we assumed that the system correlation function has little power at high frequency. Furthermore, such an assumption seems physically reasonable; the natural frequencies of the system dynamics are set by the energy splitting of the logical states and by the characteristic time scale ({\em e.g.}, the gate duration $t_0$) on which the time-dependent system Hamiltonian varies. Unfortunately, though, finding a rigorous bound on the high-frequency hybrid system correlation function is not trivial, because the hybrid Hamiltonian includes the system-bath coupling $H_{SB}$, an unbounded operator. If the bath has a low temperature, then we expect that high-frequency bath oscillators are likely to be in their ground states, but to prove a threshold theorem, we need to rule out relatively unlikely events that might foil the computation. That is not so easy to do, especially if the Hamiltonian is unbounded. So in this paper we will mostly pursue the consequences of the crude estimate eq.~(\ref{class1-sum-bound}) and other similar estimates, leaving for future work the challenge of improving the results via tighter bounds on the integral in eq.~(\ref{class1-sum}). However, we {\em can} obtain a stronger bound for the case of pure dephasing noise, discussed in Sec.~\ref{sec:diagonal}.

To prevent confusion, we remark that our ``hybrid picture'' is a rather strange concept, in that the Hamiltonian that governs evolution on the upper branch of the Keldysh diagram is different than the Hamiltonian for the lower branch. If we were using Keldysh diagrams the way they are usually used, to track the evolution of the system's density operator, this feature would be unacceptable because time evolution would not preserve the density operator's trace. For us, though, the hybrid Hamiltonian is merely a technical trick for bounding the sum of a class of diagrams, and should not be interpreted literally as the Hamiltonian of a physical system.

\subsubsection{Class 2 diagrams}

Now consider the class 2 diagrams. The earliest insertions inside the marked location on the upper and lower branches of the Keldysh diagram are not contracted with one another; rather each is contracted with an insertion at another location. Let us say that the earliest insertion at $(x,t)$ in the marked location on the upper branch is contracted with an insertion at spacetime position $(z,u)$, which could be on either the upper or lower branch, and that the earliest insertion at $(y,s)$ in the marked location on the lower branch is contracted with an insertion at $(w,v)$, which also could be on either the upper or lower branch. In principle $z$, $w$ could be the spatial labels of any two qubits in the computer, and $u$, $v$ could be any time between the initial and final time, {\em except} that the insertions at $(z,u)$ and $(w,v)$ must be {\em legal}; that is, neither can be inside the marked location on the upper branch earlier than $t$, or inside the marked location on the lower branch earlier than $s$. 

For the class 2 diagrams, let us for now imagine fixing the insertions at $(z,u)$ and at $(w,v)$ that are contracted with the earliest insertions; we will integrate over these spacetime positions later on. The simplest diagram in the class, the ``skeleton'' diagram with only two contractions, has the value (except for a phase factor that depends on the choice of branch for the insertions at $(z,u)$ and $(w,v)$)
\begin{eqnarray}
\label{skeleton-class2}
&& \langle \Psi_S^0| T^*\big(\sigma_{\beta}(y,s) \sigma_{\delta}(w,v) \sigma_{\gamma}(z,u)\sigma_{\alpha}(x,t)\big)|\Psi_S^0\rangle\nonumber\\
&& \times \langle\Psi_B^0| T^*\big(\phi_{\beta}(y,s) \phi_{\delta}(w,v)\big)|\Psi_B^0\rangle
\langle\Psi_B^0|T^*\big(\phi_{\gamma}(z,u) \phi_{\alpha}(x,t) \big)|\Psi_B^0\rangle~.
\end{eqnarray} 
where $T^*$ denotes the proper Keldysh ordering. Other diagrams in class 2 are obtained by dressing this skeleton with additional insertions and contractions in all possible legal ways. As in our discussion of the class 1 diagrams, summing all the ways to dress the skeleton transforms the interaction-picture system operators into hybrid-picture operators, yielding (up to a phase)
\begin{eqnarray}
\label{class2}
&& \langle \Psi_{SB}^0| T^*\big(\sigma^{\rm hyb}_{\beta}(y,s) \sigma^{\rm hyb}_{\delta}(w,v) \sigma^{\rm hyb}_{\gamma}(z,u)\sigma^{\rm hyb}_{\alpha}(x,t)\big)|\Psi_{SB}^0\rangle\nonumber\\
&& \times \langle\Psi_B^0| T^*\big(\phi_{\beta}(y,s) \phi_{\delta}(w,v)\big)|\Psi_B^0\rangle
\langle\Psi_B^0|T^*\big(\phi_{\gamma}(z,u) \phi_{\alpha}(x,t) \big)|\Psi_B^0\rangle~.
\end{eqnarray} 
To obtain the sum of all class 2 diagrams, we now sum over Pauli operator labels and spacetime positions, obtaining
\begin{eqnarray}
\label{class2-sum}
&&\sum \mbox{Class 2 Diagrams}\nonumber\\
&&=\sum_{x,y \in {\rm Loc}} \, \sum_{z,w \in {\rm All}'} \int_{\rm Loc}dt \int_{\rm Loc} ds \int_{{\rm All}'}du \int_{{\rm All}'} dv \sum_{\alpha,\beta, \gamma,\delta}({\rm phase})\nonumber\\
&& \times\langle \Psi_{SB}^0| T^*\big(\sigma^{\rm hyb}_{\beta}(y,s) \sigma^{\rm hyb}_{\delta}(w,v) \sigma^{\rm hyb}_{\gamma}(z,u)\sigma^{\rm hyb}_{\alpha}(x,t)\big)|\Psi_{SB}^0\rangle\nonumber\\
&& \times \langle\Psi_B^0| T^*\big(\phi_{\beta}(y,s) \phi_{\delta}(w,v)\big)|\Psi_B^0\rangle
\langle\Psi_B^0|T^*\big(\phi_{\gamma}(z,u) \phi_{\alpha}(x,t) \big)|\Psi_B^0\rangle~.
\end{eqnarray} 
Here the notation All$'$ indicates that the qubit positions $z$ and $w$ are summed over {\em both} branches of the Keldysh diagram, and that the times $u$ and $v$ are also integrated over both branches. Furthermore, it is understood that the integral over $u$ and $v$ is restricted to legal insertions (times in the upper-branch marked location earlier than $t$, and in the lower-branch marked location earlier than $s$, are excluded).

As for the class 1 diagrams, we obtain a bound on the sum of class 2 diagrams by noting that the expectation value of the product of system operators has modulus no larger than one, finding
\begin{eqnarray}
\label{class2-sum-bound}
\left|\sum\mbox{Class 2 Diagrams}\right| \le \left( 2\int_{1,{\rm Loc}}\int_{2,{\rm All}}|\bar\Delta(1,2)|\right)^2~.
\end{eqnarray}
To obtain eq.~(\ref{class2-sum-bound}) we have noted that the Keldysh ordering is irrelevant when we take the modulus of the bath two-point function, and that in the sum of the moduli of all diagrams we can extend the integral over legal insertions to an integral over all insertions to obtain an upper bound. Here the notation All indicates that the second leg of the correlation function is summed over all qubits and integrated over all times; the factor of 2 accompanies the integral $\int_{2,{\rm All}}$ because the insertions at $(z,u)$ and at $(w,v)$ can be on either one of the two branches of the Keldysh diagram.

For our upper bound on the sum of class 1 diagrams, both legs of the bath's two-point function are integrated over the marked location, while in the upper bound on the sum of class 2 diagrams, one leg is integrated over the marked location, while the other is integrated over all qubits and all times. This distinction is not so important if the spatial and temporal correlations decay rapidly, but it can be quite important if the decay is slow, as we have already discussed in Sec.~\ref{sec:dimension}. The upper bound on the sum of class 1 diagrams is still valid, though weaker, if we extend the integral for one of the legs from the marked location to all of spacetime. Then by adding together the contributions from diagrams of both classes, we find 
\begin{equation}
\label{r-equals-one-bound}
\|\Psi_{SB}^{\rm bad}(\mathcal{I}_{r=1})\|^2 \le E + 4E^2~,
\end{equation}
where 
\begin{equation}
\label{E-defined}
E= \max_{\rm Loc}\left(\int_{1,{\rm Loc}}\int_{2,{\rm All}}|\bar\Delta(1,2)|\right)~.
\end{equation}
If $E$ is small (the typical case of interest), then the class 1 diagrams dominate, and the contribution from class 2 diagrams is higher order in $E$. We emphasize again that the integral $\int_{2,{\rm All}}$ in the definition of $E$ is confined to a single branch of the Keldysh diagram, and that the factor of 2 in eq.~(\ref{class2-sum-bound}) arises because the insertion inside the marked location can be contracted with an insertion on either branch.

\subsection{Many marked locations}
\label{subsec:many-marked}
Now we want to consider the case where there are $r$ marked locations. The perturbation $H_{SB}$ must be inserted at least once in each of the $r$ marked locations, on both the upper and lower branches of the Keldysh diagram. In order to get an upper bound on the sum of all such diagrams, we will organize the sum following the same ideas as in our discussion of the $r=1$ case. In each marked location on each branch, there must be an earliest insertion of the perturbation, and this earliest insertion is contracted with another insertion elsewhere, which could be on either branch. 

\begin{figure*}[t]
\begin{center}
\vspace{-.5cm}
\includegraphics[height=5cm,keepaspectratio]{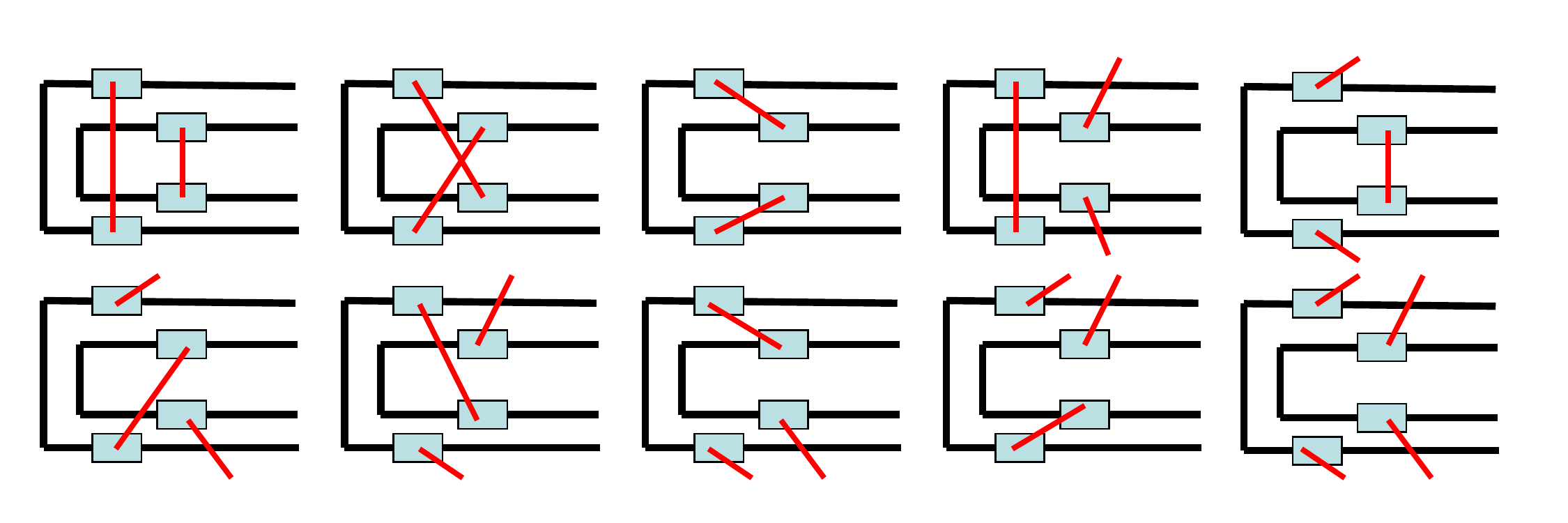}
\vspace{-.5cm} 
\end{center}
\caption{\label{fig:r-equals-two} Skeleton Keldysh diagrams contributing to $\langle \Psi_{SB}^{\rm bad}(\mathcal{I}_{r=2})|\Psi_{SB}^{\rm bad}(\mathcal{I}_{r=2})\rangle$, showing the earliest insertion of $H_{SB}$ inside each of two marked locations on both branches. There are three skeleton diagrams with two internal contractions, six skeleton diagrams with one internal contraction, and one skeleton diagram with no internal contraction, where we say that a contraction is ``internal'' if it links two earliest insertions.
         }
\end{figure*}

A skeleton graph contains a ``minimal'' set of contractions --- each contraction in the skeleton has at least one leg attached to the earliest insertion in a marked location. We distinguish two types of contractions in the skeleton: an ``internal'' contraction links two earliest insertions, and an ``external'' contraction links an earliest insertion with another legal insertion which is not an earliest insertion. The skeleton diagrams can be classified according to the number $k$ of internal contractions. If there are $r$ marked locations, and therefore all together $2r$ marked locations between the two branches, then $k$ can vary from 0 to $r$; if there are $k$ internal contractions then there are $2(r-k)$ external contractions. For $r=1$, a skeleton diagram in what we called class 1 has $k=1$ internal contractions, and a skeleton diagram in class 2 has $k=0$ internal contractions. For $r=2$, the ten distinct skeleton diagrams are shown in Fig.~\ref{fig:r-equals-two}. There are three diagrams with two internal contractions, six diagrams with one internal contraction, and one diagram with no internal contractions.

The value of a skeleton diagram, with all insertions and contractions fixed, can be expressed as a product of the expectation value of a string of Keldysh-ordered interaction-picture system operators
\begin{eqnarray}
\label{skeleton-k-internal-system}
\langle \Psi_S^0| T^*\big(
\sigma(i_1)\cdots \sigma(i_k) 
\sigma(j_1)\cdots \sigma(j_k)
\sigma(m_1)\cdots \sigma(m_{2(r-k)}) \sigma(n_1)\cdots \sigma(n_{2(r-k)})
\big)|\Psi_S^0\rangle
\end{eqnarray} 
and a product of Keldysh-ordered bath two-point functions
\begin{eqnarray}
\label{skeleton-k-internal-bath}
&&  {\rm (phase)}\times \langle\Psi_B^0| T^*\big(\phi(i_1) \phi(j_1)\big)|\Psi_B^0\rangle \cdots \langle\Psi_B^0| T^*\big(\phi(i_k) \phi(j_k)\big)|\Psi_B^0\rangle \nonumber\\
&& \times\langle\Psi_B^0|T^*\big(\phi(m_1) \phi(n_1) \big)|\Psi_B^0\rangle\cdots 
 \langle\Psi_B^0| T^*\big(\phi(m_{2(r-k)}) \phi(n_{2(r-k)}\big)|\Psi_B^0\rangle ~.
\end{eqnarray} 
Here we have attached labels $1,2,3, \dots, 2r$ to the $2r$ marked locations on the two branches, and {e.g.}, $\sigma(i_1)$ is shorthand for $\sigma_{\alpha_{i_1}}(x_{i_1},t_{i_1})$, where $(x_{i_1},t_{i_1})$ is the spacetime position of the first insertion inside marked location number $i_1$. In eqs.~(\ref{skeleton-k-internal-system}) and (\ref{skeleton-k-internal-bath}), locations $i_1$ through $i_k$ are internally contracted with locations $j_1$ through $j_k$, while the remaining earliest insertions in locations $m_1$ to $m_{2(r-k)}$ are contracted with insertions labeled $n_1$ through $n_{2(r-k)}$ which are not earliest insertions.  

When we sum up all ways to dress the skeleton with additional insertions and contractions, we obtain an expression of the same form, but with the interaction-picture system operators replaced by hybrid-picture system operators, and $|\Psi_{S}^0\rangle$ replaced by $|\Psi_{SB}^0\rangle$. Bounding the system operator expectation value by one, and summing over the Pauli operator labels, we obtain a bound on the sum of all dressed skeleton diagrams, 
\begin{eqnarray}
\left|\sum \mbox{Dressed Skeletons}\right| \le \prod_{a=1}^k |\bar\Delta(i_a,j_a)|\prod_{b=1}^{2(r-k)}|\bar\Delta(m_b,n_b)|~.
\end{eqnarray}
Now, keeping fixed the choice of which locations are internally contracted with one another, we can integrate each $i_a$, $j_a$, and $m_b$ over the specified marked location, while integrating $n_b$ over all locations on both branches. The integral is bounded above by
\begin{equation}
\label{bound-contractions-fixed}
\int \left|\sum \mbox{Dressed Skeletons}\right| \le \prod_{a=1}^k G(i_a,j_a)\prod_{b=1}^{2(r-k)} 2E(m_b)~,
\end{equation}
where 
\begin{equation}
G(i_a,j_a) = \int_{1,{\rm Loc}(i_a)}\int_{2,{\rm Loc}(j_a)}|\bar\Delta(1,2)|~,\quad 
E(m_b)= \int_{1,{\rm Loc}(m_b)}\int_{2,{\rm All}}|\bar\Delta(1,2)|~,\quad 
\end{equation}
and where the factor of 2 multiplying $E(m_b)$ results from summing $n_b$ over both branches.

Now, with the number $k$ of internal contractions still fixed, we can sum eq.~(\ref{bound-contractions-fixed}) over all the ways that $k$ contracted pairs of locations can be chosen from among $2r$ locations. We note that
\begin{equation}
\sum_{{\rm contractions}(k)}\left(\prod_{a=1}^k G(i_a,j_a) \right)\le \frac{1}{ k!}\left(\sum_{i,j=1\atop i<j}^{2r} G(i,j)\right)^k~;
\end{equation}
this inequality holds because $\left(\sum_{i,j=1}^{2r} G(i,j)\right)^k$ contains the term corresponding to each contraction $k!$ times, and also contains other nonnegative terms. Furthermore, 
\begin{equation}
\sum_{i,j=1\atop i<j}^{2r} G(i,j)\le \sum_{i=1}^{2r} E(i)~,
\end{equation}
because the expression on the right-hand-side contains all the terms on the left-hand side, plus other nonnegative terms. We conclude that 
\begin{equation}
\sum_{{\rm contractions}(k)}\int \left|\sum \mbox{Dressed Skeletons}\right| \le \frac{1}{k!} \left( 2rE\right)^k (2E)^{2(r-k)}~,
\end{equation}
with $E$ defined as in eq.~(\ref{E-defined}).

It remains to sum over $k$, the number of internal contractions:
\begin{equation}
\left| \sum \mbox{Diagrams}\right| \le \sum_{k=0}^r \left(\sum_{{\rm contractions}(k)}\int \left|\sum \mbox{Dressed Skeletons}\right| \right)\le \sum_{k=0}^r \frac{r^k}{k!} (2E)^{2r-k}~.
\end{equation}
Therefore, if we assume that $2E \le 1$,
\begin{equation}
\label{diagram-bound-hooray}
\||\Psi_{SB}^{\rm bad}(\mathcal{I}_{r})\rangle\|^2 \le (2E)^r \sum_{k=0}^r \frac{r^k}{k!} (2E)^{r-k}
\le (2E)^r \sum_{k=0}^\infty \frac{r^k}{k!} = (2eE)^r = \varepsilon^{2r}~,
\end{equation}
where
\begin{equation}
\varepsilon ~=~ \sqrt{2eE}~ \approx ~2.34 \sqrt{E}~.
\end{equation}
Thus we have derived eq.~(\ref{result}). We note that eq.~(\ref{diagram-bound-hooray}) also applies for $r=1$, and in that case is weaker than the upper bound $\||\Psi_{SB}^{\rm bad}(\mathcal{I}_{r=1})\rangle\|^2\le E+4E^2 = E(1+4E)\le 3E$ found in eq.~(\ref{r-equals-one-bound}), assuming $E \le 1/2$.

\section{Generalizations}
\label{sec:generalizations}

\subsection{Initial state of the bath}
\label{subsec:initial-state}
In our analysis, we have found it convenient to assume that the initial state of the bath is a pure state, but the analysis also applies if the bath starts out in a mixed state. Actually, we can ``purify'' a mixed state $\rho_B^0$ of the bath by introducing a fictitious reference system $R$, and choosing the pure state $|\Psi_{BR}^0\rangle$ of $BR$ so that 
\begin{equation}
\rho_B^0 ={\rm tr}_R \left(|\Psi_{BR}^0\rangle\langle \Psi_{BR}^0|\right)~.
\end{equation}
Our previous analysis then applies, if we consider $BR$ to be an ``extended'' bath, such that the system interacts with only the subsystem $B$ of the extended bath.

However, the state $|\Psi_{BR}^0\rangle$ is not arbitrary; for our argument to apply it must be chosen so that the interaction-picture free field $\phi_\alpha(x,t)$ has Gaussian statistics in this state. For this it suffices for $|\Psi_{BR}^0\rangle$ to be an {\em undisplaced Gaussian squeezed state}. If we consider the reference system $R$, like the bath $B$, to be a system of uncoupled oscillators, then an undisplaced Gaussian squeezed state is obtained by applying a unitary transformation $V$ to the oscillator ground state $|0_B,0_R\rangle$, where the action of $V$ on the annihilation operators is homogeneous and linear:
\begin{equation}
\label{bogolubov}
V^{-1} a_k V = \sum_j M_{kj} a_j + \sum_{j}N_{kj} a_j^\dagger~;
\end{equation}
here the set $\{a_k\}$ includes annihilation operators for both the $B$ oscillators and the $R$ oscillators, and the matrices $M$ and $N$ obey constraints that ensure preservation of the commutation relations. $V$ satisfies eq.~(\ref{bogolubov}) if its logarithm is strictly quadratic in creation and annihilation operators, with no linear term. A special case is the thermal state of the bath, whose purification can be written as 
\begin{eqnarray}
\exp\left( \sum_k r_k \left(a_{B,k}^\dagger a_{R,k}^\dagger - a_{B,k} a_{R,k}\right)\right)|0_B,0_R\rangle 
=\bigotimes_k\left( \sqrt{1-\gamma_k^2}\sum_{n_k=0}^\infty\gamma_k^{n_k} |\left(n_k\right)_B,\left(n_k\right)_R\rangle\right)~,
\end{eqnarray}
where $\gamma_k^2=\tanh^2 r_k=e^{-\beta\omega_k}$ and $\beta$ is the inverse temperature. But our arguments apply to any Gaussian state $V|0_B,0_R\rangle$, since the action of $V$ in eq.~(\ref{bogolubov}) maps free fields to new free fields that still satisfy Wick's theorem and have mean zero.

\subsection{Measurement and entropy removal}
For fault-tolerant computing to work, there must be a mechanism for flushing the entropy introduced by noise. In the scheme we have analyzed, entropy is removed from the computer because error-correction gadgets use a supply of fresh ancilla qubits that are discarded after use. It has been understood that the initial state $|\Psi_S^0\rangle$ of the system includes all of the ancilla qubits that will be needed during the full course of the computation. But to model the actual situation, in which ancilla qubits are prepared as needed just before being used, we imagine that ancilla qubits are perfectly isolated from the bath until ``opened'' at the onset of the gadget in which they participate. Similarly, we imagine that the measurements of all ancilla qubits are delayed until the very end of the computation, but that these qubits are ``closed'' (their coupling to the bath is turned off) at the conclusion of the gadget in which they participate. With these stipulations, our noise model is equivalent to one in which ancilla qubits are repeatedly measured, reset, and reused. 

We model the noisy preparation of an ancilla qubit as an ideal preparation followed by interaction with the bath for a specified duration. Since the state of the bath may evolve during the computation, the noise in the preparation may also depend on when the qubit is prepared. Still, we are taking it for granted that ``pretty good'' fresh ancillas can be prepared at any time, or equivalently that qubits can be effectively erased at any time. Implicitly, we have adopted a ``two-reservoir'' hypothesis. One reservoir, which we have called the ``bath,'' interacts with the system qubits, causing noise. The other reservoir is the entropy ``sink,'' which carries away heat each time a qubit is erased. In our model, the bath and the sink are uncoupled, and the sink has infinite heat capacity --- it never heats up no matter how many qubit erasures occur.

Because the bath interacts with the system, in principle it might be driven far from its initial state in a manner that depends on the ideal computation being simulated. Our arguments have shown that, at least if the bath is a system of uncoupled oscillators and its initial state is Gaussian, the bath will not be pushed to a highly adversarial state that overpowers our efforts to make the computation robust. 
One wonders how that conclusion could be altered if we relax the two-reservoir hypothesis by coupling the sink and the bath, or by eliminating the sink entirely. For example, we could attempt to model measurement and erasure more realistically by including entropy flow from the system to the bath. In that case, a bath of unbounded heat capacity would be needed to remove entropy from a noisy computation of unbounded size, and our modeling would need to incorporate a mechanism for equilibration of the bath. The goal would be to specify conditions under which the entropy flow from system to bath can be maintained well enough to support scalable quantum computation. For now, we put aside this ambitious project as an open problem for future consideration.

\subsection{Postselection}

Some fault-tolerant gadgets include {\em postselection}. For example, a gadget might consume a disposable piece of ``quantum software,'' an encoded ancilla state that is prepared offline and verified before coming into contact with the encoded data processed by the computation. The verification procedure includes measurements that check the accuracy of the preparation of the software, and the software is accepted only if the measurements have suitable outcomes; otherwise the software is rejected and the preparation is repeated. Therefore, estimates of the reliability of gadgets are conditioned on acceptance of the software, which is said to have a ``postselected'' state. 

Some fault-tolerant protocols, such as those analyzed in \cite{knill,AGP2,Reichardt06} make ``extreme'' use of postselection, meaning that the software is usually rejected and the preparation is typically repeated many times before it finally succeeds. For such protocols, noise with adversarial correlations can be a formidable foe, since the adversary is  empowered to enhance the probability of acceptance for atypical fault paths that are especially damaging. Thus, the threshold estimates based on extreme postselection proved in \cite{AGP2,Reichardt06} apply for independent noise but not for local stochastic noise. But other protocols, such as those analyzed in \cite{AGP,Aliferis07,Aliferis06c}, make only ``modest'' use of postselection, meaning that software is accepted with reasonably high probability. For such protocols, a gadget's failure rate, conditioned on acceptance of the software, can be easily estimated using the Bayes rule, even for the case of local stochastic noise. 

The threshold estimate for local noise whose proof is sketched in Appendix A applies to a protocol with no postselection at all. For local noise, as for local stochastic noise, we do not know how to extend this proof to a protocol with extreme postselection. But it {\em can} be extended to a protocol with modest postselection. This observation is useful, because threshold estimates based on protocols with modest postselection are typically higher than estimates based on protocols without postselection.

Before considering the case of local coherent noise, we recall how protocols with modest postselection can be analyzed for the case of local stochastic noise \cite{AGP}; to be concrete, we will discuss the case where  the level-1 gadgets are based on a quantum error-correcting code that corrects one error in a block. A properly designed gadget processes encoded data correctly if the software is accepted and the gadget contains no more than one fault. Therefore, the {\em joint} probability $P_{\rm joint}$ of acceptance of the software and failure of the gadget is bounded above by $B\varepsilon^2+D\varepsilon^3$ for local stochastic noise with strength $\varepsilon$, where $B$ is the number of malignant pairs of locations in the gadget where faults can cause failure (assuming the software is accepted), and $D$ is the total number of sets of three locations in the gadget. On the other hand, the software will surely be accepted if there are no faults in the software preparation circuit, so the probability of acceptance $P_{\rm accept}$ is bounded below by $1-C\varepsilon$, where $C$ is the total number of locations in the circuit for software preparation and verification. Using the Bayes rule, we obtain an upper bound on the probability $P_{\rm conditional}$ of failure conditioned on acceptance:
\begin{eqnarray}
\label{P-conditional}
P_{\rm conditional}=\frac{P_{\rm joint}}{P_{\rm accept} }\le \frac{B\varepsilon^2+D\varepsilon^3}{1-C\varepsilon}\le \varepsilon^2/\varepsilon_0= \varepsilon^{(1)} ~,
\end{eqnarray}
where
\begin{eqnarray}
\varepsilon_0^{-1}=\frac{1}{2}(B+C)\left(1+\sqrt{1+4D/(B+C)^2}\right)~
\end{eqnarray}
is determined by solving the equation $(B\varepsilon_0^2 + D\varepsilon_0^3)/(1-C\varepsilon_0)=\varepsilon_0$.
This argument gives a useful result if our lower bound on $P_{\rm accept}$ is not too small. In practice, it is often the case that $C\ll B$ and therefore $C\varepsilon_0\ll 1$, so that the ``postselection correction'' arising from division by $P_{\rm accept}$ is a small effect.

There is another way to describe this estimate that is more readily generalized to the case of local coherent noise, and also clarifies why the estimate applies to adversarial local stochastic noise. Imagine that $n$ software preparation and verification attempts are executed in parallel, where we label the attempts by an index $i=1,2,3,\dots, n$, and suppose for the moment that the noise is uncorrelated. Now we distinguish $n+1$ possible ways for the gadget to be bad, depending on which preparation attempt (if any) is the first to be accepted. If ancilla 1 is accepted, then the gadget fails with probability $P_{\rm joint}$. But ancilla 1 is rejected with probability $P_{\rm reject}=1-P_{\rm accept}$, so the probability that ancilla 1 is rejected, ancilla 2 is accepted, and the gadget fails is $P_{\rm reject}P_{\rm joint}$. Similarly, the probability that ancilla $m$ is the first to be accepted and the gadget fails is $P_{\rm reject}^{m-1}P_{\rm joint}$, and the probability that all $n$ ancillas are rejected is $P_{\rm reject}^n$. Summing the probability of all failure scenarios, we find
\begin{eqnarray}
P_{\rm fail}= P_{\rm joint}\left(\sum_{m=1}^n P_{\rm reject}^{m-1}\right) + P_{\rm reject}^n = \frac{P_{\rm joint}}{P_{\rm accept} }\left(1-P_{\rm reject}^n\right) +P_{\rm reject}^n
=\frac{P_{\rm joint}}{P_{\rm accept} } + P_{\rm reject}^n\left(1-\frac{P_{\rm joint}}{P_{\rm accept}}\right)~.
\end{eqnarray}
In the limit $n\to\infty$, we recover the estimate eq.~(\ref{P-conditional}), and even for $n=2$ we have $P_{\rm fail}=O(\varepsilon^2)$. Furthermore, the upper bound on $P_{\rm fail}$ obtained from upper bounds on $P_{\rm joint}$ and $P_{\rm reject}$ applies not only to independent noise but also to correlated local stochastic noise --- it can be regarded as an estimate of the effective noise strength $\varepsilon^{(1)}$ after one level reduction step. For local stochastic noise, we sum over all failure scenarios at each of $r$ marked locations, and conclude that the probability that all $r$ locations are bad is bounded above by $\left(\varepsilon^{(1)}\right)^r$.

We can also apply this strategy of summing over all failure scenarios in the case of local coherent noise. First we note that, to preserve the framework assumed in Sec.~\ref{sec:derivation}, we may imagine that all measurements in verification steps are postponed until the end of the computation. In the actual circuit, the ``verification qubits'' are measured inside gadgets, and then subsequent operations are conditioned on the classical measurement outcomes. To model this circuit in the framework where all measurements are postponed, we suppose that a verification qubit decouples from the bath at the time when it is measured in the actual circuit, and we replace operations conditioned on measurement outcomes by noiseless quantum gates conditioned on the state of the verification qubit, after decoupling from the bath but prior to being measured. Then we can estimate $\| |\Psi_{SB}^{\rm bad}(\mathcal{I}_r^{(1)})\rangle\|^2$ by summing over $n + 1$ failure scenarios at each of the $r$ marked locations. In scenario 1, ancilla 1 is accepted and the gadget using ancilla 1 (including the preparation and verification of ancilla 1) has two or more faults. In scenario $m$, for $m=2,3, \dots, n$, the first $m-1$ ancillas are rejected, ancilla $m$ is accepted, and the gadget using ancilla $m$ has two or more faults. In scenario $n+1$, all $n$ ancillas are rejected. Since the scenarios are perfectly distinguishable, they should be summed incoherently.

Now, in order for an ancilla to be rejected, there must be at least one fault in the circuit that prepares and verifies that ancilla. Therefore, in scenario $m$, we sum coherently over all fault paths such that there is at least one fault in each of the first $m-1$ ancilla preparation/verification circuits, and at least two faults in the gadget using ancilla $m$. This sum includes all of the fault paths that contribute to the badness of the gadget under scenario $m$, but it also includes other fault paths that do not contribute to scenario $m$. However, since the scenarios are distinguishable, there is no harm in including these additional unwanted scenarios, if our goal is to obtain an upper bound on $\| |\Psi_{SB}^{\rm bad}(\mathcal{I}_r^{(1)})\rangle\|^2$. This coherent sum for each scenario can be estimated by the method described in Appendix A. One finds that, for gadgets such that the ``postselection correction'' to $\varepsilon^{(1)}$ is small in the case of local stochastic noise, the correction is small for local coherent noise as well. 

In \cite{Aliferis06c}, the lower bound on the accuracy threshold $\varepsilon_0 \ge 1.94 \times 10^{-4}$ was established for local stochastic noise, based on a protocol with modest use of postselection. Though we have not done the calculation in detail, we expect that a similar estimate $\varepsilon_0\sim 10^{-4}$, based on the same protocol, also applies to the threshold noise strength for local coherent noise. (The argument in \cite{fibonacci} achieves a higher threshold estimate for local stochastic noise, but uses a different method that is less easily adapted to the case of local coherent noise.)

Of course, for the case of Gaussian noise, if gadgets include multiple parallel attempts to prepare and verify software, then all of these attempts should be included in the integral $\int_{2,{\rm All}}$ in our estimate of the noise strength in eq.~(\ref{result}).

\subsection{Other considerations}
It would be desirable to extend the derivation of our threshold result in several other directions. One possible approach is to allow the bath fluctuations to be weakly non-Gaussian by including small anharmonic corrections in the bath Hamiltonian $H_B$. But, though the effects of bath self-interactions can be analyzed perturbatively by standard methods, obtaining useful rigorous results summed to all orders of perturbation theory is not simple. Another worthy goal, already emphasized at the end of Sec.~\ref{sec:implications}, is to formulate a threshold condition less sensitive to the high-frequency fluctuations of the bath; {i.e.}, to noise with a frequency large compared to the natural frequencies of the ideal system Hamiltonian $H_S$. In principle this might be done by ``integrating out'' high-frequency noise, obtaining an effective noise model with a lower frequency cutoff that faithfully reproduces the impact of the noise on the simulated computation. Making such an analysis rigorous is another challenging open problem. In the next Section, though, we will discuss one special case in which an improved threshold estimate less sensitive to high-frequency noise can be achieved.

\section{Diagonal Gaussian noise}
\label{sec:diagonal}

As we discussed in Sec.~\ref{subsubsec:class1}, our general arguments do not place any constraints on the frequency spectrum of the ``hybrid-picture'' system operators. Therefore, we were forced to take the modulus of the bath two-point function in our estimate of the noise strength $\varepsilon$. And as a result, our estimate has a sensitivity to high-frequency bath fluctuations that seems rather artificial. 

There is at least one case where we have much better analytic control over the time-dependence of the system operators, allowing us to obtain a better estimate of the noise strength that has milder sensitivity to high-frequency noise. That is the case of pure dephasing noise, which we will discuss now.

In this noise model, the bath couples only to the $z$-components of the qubits, so that the system-bath Hamiltonian is 
\begin{equation}
\label{diagonal-system-bath}
H_{SB}=\sum_x \sigma_z(x)\otimes \tilde\phi(x,t)~,
\end{equation}
where $\tilde\phi(x,t)$ is a Gaussian bath variable with mean zero. To further simplify the discussion (whose purpose is merely illustrative anyway), we will also assume there are no multi-qubit correlations in the noise (even though this might not be an accurate description of the noise in multi-qubit gates). That is, we assume $\langle\phi(x,t)\phi(y,s)\rangle=0$ for $x\ne y$, so that in effect each qubit is coupled to its own independent bath.

A scheme for fault-tolerant quantum computation customized for highly-biased noise dominated by dephasing was formulated in \cite{AP-bias} and further discussed in \cite{brito}. In this scheme, all gates are teleported. Furthermore the only fundamental operations used are single-qubit preparations, single-qubit measurements in the $\sigma_x$-eigenstate basis, and two-qubit controlled-phase ({\sc cphase}) gates. A {\sc cphase} gate is diagonal in the computational ({\em i.e.}, $\sigma_z$-eigenstate) basis, with eigenvalues $(1,1,1,-1)$. Thus it can be realized by a time-dependent two-qubit system Hamiltonian that is also diagonal:
\begin{equation}
H_S= f(t)\left( \sigma_z\otimes \sigma_z - \sigma_z\otimes I - I\otimes\sigma_z\right)~,
\end{equation}
where $\int dt f(t) = \pi/4$. This diagonal system Hamiltonian commutes with the system-bath Hamiltonian eq.~(\ref{diagonal-system-bath}), whose action on the system qubits is diagonal. As in previous Sections, we model a noisy qubit preparation as an ideal preparation followed by interaction with the oscillator bath, and we model a noisy measurement as interaction with the bath followed by an ideal measurement.

We can analyze the effect of the noise on the computation using interaction-picture perturbation theory, and in fact we can estimate a {\em probability} (rather than an amplitude) for the outcome of a qubit measurement to differ from the measurement outcome in the ideal computation. For each qubit, we distinguish between {\em good} diagrams, in which the perturbation $H_{SB}$ is inserted an {\em even} number of times in between the (ideal) qubit preparation and the (ideal) qubit measurement, and the {\em bad} diagrams, in which the perturbation is inserted an {\em odd} number of times in between the preparation and the measurement. Because $\sigma_z$ commutes with the ideal system Hamiltonian $H_S$, and because $\sigma_z^2=I$, in all good diagrams the outcome of the final $\sigma_x$ measurement agrees with the result in the ideal quantum circuit, while in bad diagrams the measurement outcome is flipped. 

Furthermore, the good and the bad part of the system-bath state $|\Psi_{SB}\rangle$ are mutually orthogonal. To see this, imagine evaluating the inner product $\langle\Psi^{\rm good}_{SB}|\Psi^{\rm bad}_{SB}\rangle$ between the good and bad parts of the state for a single qubit and its associated bath. Since $|\Psi^{\rm bad}_{SB}\rangle$ has an odd number of perturbation insertions and $|\Psi^{\rm good}_{SB}\rangle$ has an even number, each Keldysh diagram contributing to $\langle\Psi^{\rm good}_{SB}|\Psi^{\rm bad}_{SB}\rangle$ is proportional to the expectation value of a product of an odd number of interaction-picture bath fields. All such diagrams vanish, since $\phi(x,t)$ is Gaussian with mean zero. Because the good and bad parts of $|\Psi_{SB}\rangle$ are perfectly distinguishable, we can regard $\langle\Psi^{\rm bad}_{SB}|\Psi^{\rm bad}_{SB}\rangle$ as the {\em probability} of error in the final qubit measurement. 

\begin{figure*}[t]
\begin{center}
\vspace{-.5cm}
\includegraphics[height=3cm,keepaspectratio]{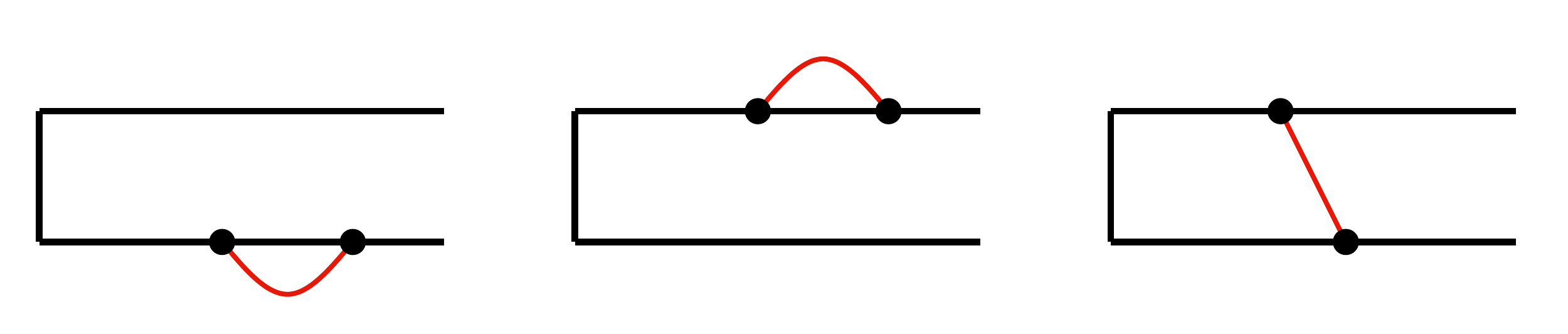}
\vspace{-.5cm} 
\end{center}
\caption{\label{fig:dephasing} The three connected Keldysh diagrams for a single qubit subject to Gaussian dephasing noise. The first two diagrams are ``good'' because $\sigma_z$ is inserted an even number of times on each branch, and the third diagram is ``bad'' because $\sigma_z$ is inserted an odd number of times on each branch.
         }
\end{figure*}

Let us compute this probability. The sum of all Keldysh diagrams (both good and bad) contributing to $\langle\Psi_{SB}|\Psi_{SB}\rangle$ for a single qubit is the exponential of the sum of ``connected'' diagrams. There are three connected diagrams, shown in Fig.~\ref{fig:dephasing}. Thus
\begin{equation}
\label{diagonal-sb-norm}
1= \langle\Psi_{SB}|\Psi_{SB}\rangle = \exp\left( C_U + C_L + D\right)~;
\end{equation}
here, 
\begin{equation}
C_U=-\int_{t>s} dt ~ds ~\langle \phi(t)\phi(s)\rangle~
\end{equation}
is the connected diagram in which two insertions on the upper Keldysh branch are contracted, 
\begin{equation}
C_L=-\int_{t<s} dt ~ds ~\langle \phi(t)\phi(s)\rangle~
\end{equation}
is the connected diagram in which two insertions on the lower branch are contracted, and
\begin{equation}
D=\int dt ~ds ~\langle \phi(t)\phi(s)\rangle= -(C_U+C_L)~.
\end{equation}
is the connected diagram in which an insertion on the upper branch is contracted with an insertion on the lower branch. (In all three diagrams, the factor due to the expectation value of the product of system operators is simply 1.) When eq.~(\ref{diagonal-sb-norm}) is expanded in powers of $D$, terms with an odd number of powers contribute to $\langle\Psi^{\rm bad}_{SB}|\Psi^{\rm bad}_{SB}\rangle$, because $H_{SB}$ is inserted an odd number of times on each branch, and terms with an even number of powers contribute to $\langle\Psi^{\rm good}_{SB}|\Psi^{\rm good}_{SB}\rangle$, because $H_{SB}$ is inserted an even number of times on each branch.  Thus,
\begin{eqnarray}
&&{P}^{\rm bad}\equiv\langle \Psi_{SB}^{\rm bad}|\Psi_{SB}^{\rm bad}\rangle=e^{-D} \sinh D~, \nonumber\\
&&{P}^{\rm good}\equiv\langle \Psi_{SB}^{\rm good}|\Psi_{SB}^{\rm good}\rangle=e^{-D} \cosh D~.
\end{eqnarray}
If $T$ is the elapsed time between the preparation and measurement of the qubit, then 
\begin{equation}
D=\int_{-\infty}^{\infty}\frac{d\omega}{2\pi}\tilde\Delta(\omega)\frac{4\sin^2(\omega T/2)}{\omega^2}~.
\end{equation}
where 
\begin{equation}
\Delta(t-s) \equiv \langle \phi(t)\phi(s)\rangle =\int_{-\infty}^\infty \frac{d\omega}{2\pi}e^{-i\omega(t-s)}\tilde\Delta(\omega)~
\end{equation}
(we have assumed the noise is stationary). 

In the case of zero-temperature Ohmic noise, with $\tilde\Delta(\omega)$ given by eq.~(\ref{ohmic-spectrum}), we find that 
\begin{equation}
D= -\int_0^T dt \int_0^T~ds ~\frac{A}{(t-s -i\tau_c)^2}= A~\ln \left(\frac{T^2+\tau_c^2}{\tau_c^2}\right)\approx 2A~\ln(T/\tau_c)~.
\end{equation}
Thus the quantity $D$ (an upper bound on the probability $P^{\rm bad}$ of a measurement error) has only a mild logarithmic sensitivity to the ultraviolet cutoff $\tau_c^{-1}$, in contrast to the power dependence on the cutoff found in eq.~(\ref{ohmic-integral}). This improvement occurs because $D$ is found by integrating the bath two-point function $\Delta(t)$, rather than its modulus $|\Delta(t)|$, which can be justified because the perturbation $H_{SB}$ commutes with the ideal system Hamiltonian $H_S$. Even this logarithmic dependence on $\tau_c$ may be spurious; it arises because we have assumed that the ideal qubit preparation (at time $t=0$) and qubit measurement (at time $t=T$) are instantaneous. The divergence would be softened further if we used a smoother model of preparation and measurement. 

Perhaps the logarithmic dependence of the error probability on the elapsed time $T$ should not be taken too seriously; it applies only if the noise spectrum is Ohmic down to a frequency of order $T^{-1}$. Let us nevertheless pursue the implications of this behavior. The crux of the scheme formulated in \cite{AP-bias} is a teleported logical {\sc cnot} gate protected against dephasing by an $n$-qubit repetition code (where $n$ is odd). This {\sc cnot} gadget contains four logical measurements, each of which is decoded by a majority vote. Furthermore, for each qubit, there are at most $3n$ time steps (each of duration $t_0$) in between the preparation and measurement of the qubit, where a {\sc cphase} gate acts on the qubit in each step. Therefore, the probability $\varepsilon_{\rm CNOT}$ of an encoded error in this {\sc cnot} gadget can be bounded as
\begin{equation}
\label{cnot-error}
\varepsilon_{\rm CNOT} \le 4{n\choose \frac{n+1}{2}}\left(P^{\rm bad}\right)^{\frac{n+1}{2}}~,
\end{equation}
where
\begin{equation}
P^{\rm bad} ~\le ~D ~\le~ 2A \ln\big((3n+2)t_0/\tau_c\big)~
\end{equation}
(here we have allowed the noise to act for a time $t_0$ during each {\sc cphase} gate and also during the initial preparation and final measurement). Hence the logical {\sc cnot} gate is well protected if $A$ is small and $t_0/\tau_c$ is not too large.  

While the underlying noise model is Gaussian dephasing noise, under our assumptions the effective noise model for the {\sc cnot} gadgets is independent stochastic noise. Just to be specific, suppose that $A =10^{-3}$ and $t_0/\tau_c= 10^3$.  Then, for code length $n=9$, eq.~(\ref{cnot-error}) yields $\varepsilon_{\rm CNOT} < 1.85 \times 10^{-6}$. This {\sc cnot} error rate is well below the accuracy threshold for the local stochastic noise model, indicating that these logical {\sc cnot} gates are adequate for scalable quantum computing. 

\section{Conclusions}
\label{sec:conclusions}
The quantum accuracy threshold theorem indicates that scalable quantum computing is feasible in principle. But will fault-tolerant quantum computation {\em really} work? One concern is that the noise models assumed by theorists are highly idealized, at best crude approximations to the noise in actual devices. In formulating these models, one desires on the one hand to capture essential features of realistic noise, but on the other hand to allow a succinct and elegant analysis of the computation's reliability.

Seeking an appropriate balance between these two desiderata, we have proved in this paper a new version of the threshold theorem that applies to Gaussian quantum noise, which is physically well motivated and analytically tractable. Our result shows that quantum computing is scalable if the noise power spectrum obeys a certain condition. Compared to previous results regarding the effectiveness of fault-tolerant methods against non-Markovian noise \cite{terhal,AGP,AKP}, our threshold condition has two advantages: it is expressed in terms of experimentally observable features of the noise, and it is less sensitive to high-frequency noise. 

As mentioned in Sec.~\ref{sec:generalizations}, it might be useful to extend our results by relaxing the noise model in several ways, for example by including weak non-Gaussian corrections to the bath fluctuations, or by modeling more realistically the dissipative flow of heat from system to bath. It should also be possible to make further improvements in the sensitivity of the threshold condition to high-frequency noise; however, an improved condition would be likely to depend on the details of the frequency spectrum of the ideal system dynamics, and deriving it would require a more complicated analysis. 

Experimenters tend to worry less about high-frequency noise than about low-frequency noise, particularly $1/f$ dephasing noise.  We anticipate that low frequency noise in quantum gates can be suppressed substantially through clever design of pulse sequences, leaving weak residual noise to be tamed via the fault-tolerant methods we have studied here. Joining pulse shaping methods with fault-tolerant circuit construction will be a fruitful topic for future research. 

\acknowledgments
We thank Panos Aliferis, Matt Hastings, Alexei Kitaev, Eduardo Novais, and Gil Refael for useful discussions. This research is supported in part by DoE under Grant No. DE-FG03-92-ER40701, NSF under Grant No. PHY-0456720, NSA under ARO Contract No. W911NF-05-1-0294, and by the Gordon and Betty Moore Foundation.

\bibliographystyle{unsrt}

\appendix

\section{Threshold theorem for local noise}
\label{sec:appendix}

Here we will briefly sketch the proof of the quantum accuracy threshold theorem for local noise, following the argument in \cite{AGP}. 

We assume that the joint evolution of the quantum computer (the system $S$) and its environment (the bath $B$) is governed by the Hamiltonian $H=H_S+H_B +H_{SB}$, where the perturbation $H_{SB}$ is responsible for the deviation of the system from its ideal evolution. Although this framework can be generalized (as discussed in Sec.~\ref{sec:generalizations}), we also assume that the system qubits are initialized ideally in the pure state $|\Psi_S^0\rangle$ before the Hamiltonian evolution begins and measured ideally after it ends. Furthermore, the initial state of the bath is the pure state $|\Psi_B^0\rangle$. Just before the ideal measurements are performed on the system qubits, the joint state of the system and bath is $|\Psi_{SB}\rangle=U_{SB}|\Psi_{SB}^0\rangle$, where $U_{SB}$ is the time-evolution operator determined by $H$, and $|\Psi_{SB}^0\rangle=|\Psi_S^0\rangle \otimes |\Psi_B^0\rangle$. We obtain a fault-path expansion for $|\Psi_{SB}\rangle$ by expanding $U_{SB}$ as a power series in $H_{SB}$, and for each term in this expansion we declare a level-0 circuit location to be bad if $H_{SB}$ acts nontrivially somewhere within that location. 

For any specified set $\mathcal{I}_r$ of $r$ locations in the circuit,  we denote by $|\Psi_{SB}^{\rm bad}(\mathcal{I}_r)\rangle$ the sum of all the terms in the fault-path expansion of $|\Psi_{SB}\rangle$ such that all of these $r$ locations are bad. The noise is local with strength $\varepsilon$ if
\begin{equation}
\label{local-noise-strength-again}
\| |\Psi_{SB}^{\rm bad}(\mathcal{I}_r)\rangle\| \le \varepsilon^r~.
\end{equation}
Our objective is to show that scalable quantum computing is possible provided that $\varepsilon < \varepsilon_0$, where $\varepsilon_0$ is (a lower bound on) the accuracy threshold.

Suppose that a universal set of fault-tolerant level-1 gadgets can be constructed such that a 1-gadget containing fewer than $s$ faulty level-0 gates simulates the corresponding ideal 0-gate correctly. We can estimate the effective noise strength for a level-1 simulation using the following observation: Consider a set $\mathcal{I}$ of level-0 locations in a quantum circuit. Then the sum of all fault paths such that at least $s$ of the locations in the set $\mathcal{I}$ are faulty can be expressed as
\begin{eqnarray}
\label{inclusion-exclusion}
|\Psi_{SB}(\ge s {\rm ~faults ~in~}\mathcal{I})\rangle =\sum_{\ell=s}^{|\mathcal{I}|}(-1)^{\ell-s}{\ell-1\choose s-1}
\left(\sum_{\mathcal{I}_\ell\subseteq \mathcal{I}}|\Psi_{SB}^{\rm bad}(\mathcal{I}_\ell)\rangle\right)~,
\end{eqnarray}
where $\sum_{\mathcal{I}_\ell}$ denotes the sum over all subsets of $\mathcal{I}$ that contain $\ell$ elements. Eq.~(\ref{inclusion-exclusion}) follows from  the ``inclusion-exclusion principle'' of combinatorics. For example, in the case $s{=}1$ it becomes
\begin{eqnarray}
&&|\Psi_{SB}(\ge 1 {\rm ~fault~in~}\mathcal{I})\rangle \nonumber\\
&&=
\sum_{\mathcal{I}_1\subseteq \mathcal{I}}|\Psi_{SB}^{\rm bad}(\mathcal{I}_1)\rangle
-\sum_{\mathcal{I}_2\subseteq \mathcal{I}}|\Psi_{SB}^{\rm bad}(\mathcal{I}_2)\rangle
+\sum_{\mathcal{I}_3\subseteq \mathcal{I}}|\Psi_{SB}^{\rm bad}(\mathcal{I}_3)\rangle
-\sum_{\mathcal{I}_4\subseteq \mathcal{I}}|\Psi_{SB}^{\rm bad}(\mathcal{I}_4)\rangle + \cdots~,
\end{eqnarray}
whose origin is easy to understand. The first term counts correctly each fault path with exactly one fault in $\mathcal{I}$, but it double-counts each fault path with exactly two faults, and this over-counting is corrected by the second term. The first term counts three times each fault path with exactly three faults, and the second term subtracts these fault paths ${3}\choose{2}$ times; this under-counting is corrected by the third term. And so on.

The norm of the left-hand side of eq.~(\ref{inclusion-exclusion}) is bounded above by the sum of the norms of the terms on the right hand side. Using $\| \Psi_{SB}^{\rm bad}(\mathcal{I}_\ell)\rangle\| \le \varepsilon^\ell$ and denoting $|\mathcal{I}|=A$ we find  
\begin{eqnarray}
\label{s-fault-bound}
\big\||\Psi_{SB}(\ge s {\rm ~faults~in~}A{\rm ~locations})\rangle\big\| 
&&
\le \sum_{\ell=s}^{A}{\ell-1\choose s-1}{A\choose \ell}\varepsilon^\ell 
={A\choose s}\varepsilon^s\sum_{\ell=s}^A {A-\ell\choose \ell-s}\varepsilon^{\ell-s}\nonumber\\
&&
\le {A\choose s}\varepsilon^s\sum_{t=0}^\infty \frac{(A-s)^t\varepsilon^t}{t!}
= {A\choose s}e^{(A-s)\varepsilon} \varepsilon^s 
\le \zeta {A\choose s}\varepsilon^s~;
\end{eqnarray}
here $\zeta$ is a constant satisfying $\zeta \ge e^{(A-s)\varepsilon}$ for values of $\varepsilon$ in some specified range of interest, which typically can be chosen such that $\zeta$ is close to 1. 

In a level-1 simulation of a quantum circuit, let us say that a level-1 gadget is ``bad'' if it contains $s$ or more bad level-0 gates, and let $\mathcal{I}_r^{(1)}$ denote a set of $r$ specified level-1 gadgets. We assume, for now, that these level-1 gadgets are nonoverlapping; {\em i.e.}, that no 0-gate is contained in two different 1-gadgets. We denote by $|\Psi_{SB}^{\rm bad}(\mathcal{I}_r^{(1)})\rangle$ the sum of all terms in the perturbation expansion of $|\Psi_{SB}\rangle$ such that all of the $r$ 1-gadgets in $\mathcal{I}_r^{(1)}$ are bad. By performing an ``inclusion-exclusion'' sum independently inside each 1-gadget, we find that
\begin{eqnarray}
\label{inclusion-exclusion-level1}
|\Psi_{SB}^{\rm bad}(\mathcal{I}_r^{(1)})\rangle
 =\sum_{\ell_1=s}^{|\mathcal{I}(1)|}(-1)^{\ell_1-s}{\ell_1-1\choose s-1}
\cdots
\sum_{\ell_r=s}^{|\mathcal{I}(r)|}(-1)^{\ell_r-s}{\ell_r-1\choose s-1}\nonumber\\
\left(\sum_{\mathcal{I}(1)_{\ell_1}\subseteq \mathcal{I}(1)}\cdots
\sum_{\mathcal{I}(r)_{\ell_r}\subseteq \mathcal{I}(r)}
|\Psi_{SB}^{\rm bad}(\mathcal{I}(1)_{\ell_1}\cup\dots\cup \mathcal{I}(r)_{\ell_r})\rangle\right)~,
\end{eqnarray}
where $\mathcal{I}(j)$ denotes the set of 0-gates inside the 1-gadget $j$, for $j\in \{1,2,3,\dots, r\}$, and $\sum_{\mathcal{I}(j)_{\ell_j}}$ denotes the sum over all subsets of $\mathcal{I}(j)$ that contain $\ell_j$ elements. 

The local noise condition $\| \Psi_{SB}^{\rm bad}(\mathcal{I}_\ell)\rangle\| \le \varepsilon^\ell$ implies that
\begin{eqnarray}
\label{local-noise-level1}
\big\||\Psi_{SB}^{\rm bad}(\mathcal{I}(1)_{\ell_1}\cup\dots\cup \mathcal{I}(r)_{\ell_r})\rangle\big\|\le \prod_{j=1}^r\varepsilon^{\ell_j}~.
\end{eqnarray}
As above, we can bound the norm of the left-hand side of eq.~(\ref{inclusion-exclusion-level1}) by the sum of the norms of the terms on the right-hand side. Using eq.~(\ref{local-noise-level1}), this upper bound factorizes into a product of $r$ sums, each of which can be bounded as in eq~(\ref{s-fault-bound}). We obtain
\begin{eqnarray}
\| |\Psi_{SB}^{\rm bad}(\mathcal{I}_r^{(1)})\rangle\| \le \prod_{j\in 1}^r\varepsilon^{(1)}_j~,
\end{eqnarray}
where 
\begin{eqnarray}
\label{noise-strength-level1}
\varepsilon^{(1)}_j= \zeta_j {A_j\choose s} \varepsilon^s~,
\end{eqnarray}
and hence
\begin{eqnarray}
\| |\Psi_{SB}^{\rm bad}(\mathcal{I}_r^{(1)})\rangle\| \le \left(\varepsilon^{(1)}\right)^r~,
\end{eqnarray}
where
\begin{eqnarray}
\varepsilon^{(1)}= \max_j \left(\varepsilon^{(1)}_j\right)~.
\end{eqnarray}
Here $A_j=|\mathcal{I}(j)|$, $\zeta_j \ge \exp\left((A_j-s)\varepsilon\right)$, and the maximum is over all 1-gadgets in the circuit. We can regard $\varepsilon^{(1)}$ as an effective noise strength for the level-1 circuit, which is conveniently expressed in the form
\begin{eqnarray}
\varepsilon^{(1)} = \varepsilon_0\left(\varepsilon/\varepsilon_0\right)^s~,
\end{eqnarray}
where 
\begin{equation}
\label{threshold-estimate-combinatoric}
\varepsilon_0 = \min \left( \zeta {A\choose s}\right)^{-1/(s-1)}~,
\end{equation}
and the minimum is over all 1-gadgets in our universal set.

Now let us say that a level-$k$ gadget is bad if it contains $s$ or more bad $(k{-}1)$-gadgets. The bound eq.~(\ref{noise-strength-level1}) on the norm of the sum $|\Psi_{SB}^{\rm bad}(\mathcal{I}_r^{(1)})\rangle$ over all fault paths that are bad at level 1 is of the same form as the bound eq.~(\ref{local-noise-strength-again}) on the norm of the sum over all fault paths that are bad at level 0, but with a ``renormalized'' value of the effective noise strength. This means that in a recursive simulation, in which $k$-gadgets are constructed using the same circuits as 1-gadgets, but with each 0-gate in the 1-gadget replaced by a ($k{-}1$)-gadget, we can use the same combinatoric argument again to estimate the effective noise strength at level $k$. That is, suppose that 
\begin{eqnarray}
\| |\Psi_{SB}^{\rm bad}(\mathcal{I}_r^{(k-1)})\rangle\| \le \left(\varepsilon^{(k{-}1)}\right)^r~,
\end{eqnarray}
where $\mathcal{I}_r^{(k-1)}$ is any specified set of $r$ $(k{-}1)$-gadgets in a level-$(k{-}1)$ simulation, and $|\Psi_{SB}^{\rm bad}(\mathcal{I}_r^{(k-1)})\rangle$ denotes the sum of all fault paths such that all $r$ of the $(k{-}1)$-gadgets in $\mathcal{I}_r^{(k-1)}$ are bad. Then we may infer that 
\begin{eqnarray}
\label{level-k-local-noise}
\| |\Psi_{SB}^{\rm bad}(\mathcal{I}_r^{(k)})\rangle\| \le \left(\varepsilon^{(k)}\right)^r~,
\end{eqnarray}
where $\mathcal{I}_r^{(k)}$ is any specified set of $r$ $k$-gadgets in a level-$k$ simulation,  $|\Psi_{SB}^{\rm bad}(\mathcal{I}_r^{(k)})\rangle$ denotes the sum of all fault paths such that all $r$ of the $k$-gadgets in $\mathcal{I}_r^{(k)}$ are bad, and
\begin{eqnarray}
\varepsilon^{(k)}/\varepsilon_0 = \left(\varepsilon^{(k{-}1)}/\varepsilon_0\right)^s= \left(\varepsilon/\varepsilon_0\right)^{s^k}.
\end{eqnarray}

The fault-path expansion of a level-$k$ simulation with all together $L$ $k$-gadgets can be expressed as 
\begin{eqnarray}
|\Psi_{SB}\rangle = |\Psi_{SB}^{\rm good}\rangle +|\Psi_{SB}^{\rm bad}\rangle~,
\end{eqnarray}
where $|\Psi_{SB}^{\rm good}\rangle$ is the sum of all fault paths such that every $k$-gadget is good, and $|\Psi_{SB}^{\rm bad}\rangle$ is the sum of all fault paths such that at least one $k$-gadget is bad. Combining the $s{=}1$ case of eq.~(\ref{s-fault-bound}) with eq.~(\ref{level-k-local-noise}), then, we see that
\begin{eqnarray}
\big\||\Psi_{SB}^{\rm bad}\rangle\big\| \le L \exp\left((L-1)\varepsilon^{(k)}\right)\varepsilon^{(k)}~,
\end{eqnarray}
which is small for $L\varepsilon^{(k)}\ll 1$. Furthermore, the arguments in \cite{AGP} show that if $|\Psi_{SB}\rangle$ is good then the probability distribution $p^{({\rm actual})}=\{p_a^{({\rm actual})}\}$ governing the measurement outcome for the {\em logical} system qubits (where $p_a$ is the probability of the measurement outcome labeled by $a$) matches exactly the probability distribution $p^{({\rm ideal})}$ for the measurement outcomes in the ideal computation. Therefore, the deviation of the $p^{({\rm actual})}$ from $p^{({\rm ideal})}$ in the noisy simulation arises only from the small bad component of $|\Psi_{SB}\rangle$; in fact in the $L^1$ norm this deviation can be bounded as
\begin{equation}
\delta= \| p^{\rm (actual)} -p^{\rm (ideal)}\|_1=\sum_a |p_a^{\rm (actual)} -p_a^{\rm (ideal)}| \le 2 \big\||\Psi_{SB}^{\rm bad}\rangle\big\|~.
\end{equation}
Therefore, for $\varepsilon < \varepsilon_0$, the noisy computation becomes highly reliable as the level $k$ of the simulation increases; thus $\varepsilon_0$ is a lower bound on the accuracy threshold for quantum computation.

In \cite{AGP}, two valuable extensions of this argument were formulated that are useful for pushing the threshold estimate $\varepsilon_0$ higher. First, the argument can be applied to simulations where successive 1-gadgets {\em overlap}, {\em i.e.}, have 0-gates in common. By allowing the gadgets to overlap, we can justify the estimate eq.~(\ref{threshold-estimate-combinatoric}) when using properly designed gadgets based on a quantum error-correcting code that can correct $s{-}1$ errors in a code block. Second, we can refine the definition of badness, so that a 1-gadget with $s$ or more faults is declared bad only if the faults occur at a ``malignant'' set of locations, {\em i.e.}, only if the 1-gadget processes encoded information incorrectly because of the faults. For example, for gadgets that can correct one error (the $s{=}2$ case), our estimate of the level-1 effective noise strength improves to 
\begin{equation}
\varepsilon^{(1)} = B\varepsilon^2 + D\varepsilon^3 \le \varepsilon^2/\varepsilon_0~,
\end{equation}
where $B$ is the number of malignant {\em pairs} of fault locations in the 1-gadget (maximized over all 1-gadgets), $D\varepsilon^3$ is a correction arising from summing contributions from fault paths with three of more faults in the 1-gadget, and
\begin{equation}
\varepsilon_0^{-1}= \frac{1}{2}B\left(1 + \sqrt{1+4D/B^2}\right)
\end{equation}
is our improved threshold estimate, found by solving the equation $B\varepsilon_0^2 + D\varepsilon_0^3=\varepsilon_0$.


\end{document}